\documentclass[10pt]{article}

\usepackage[tbtags]{amsmath}
\usepackage{amsopn,amsthm,amsfonts,amssymb}

\newcommand{\sm}[1]{\ensuremath{m_{(#1)}}}
\newcommand{\sj}[1]{\ensuremath{\mathcal{J}_{(#1)}}}
\newtheorem{theorem}{Theorem}
\newtheorem{lemma}[theorem]{Lemma}
\newtheorem{proposition}[theorem]{Proposition}

\newtheorem{property}[theorem]{Property}

\newtheorem{corollary}[theorem]{Corollary}

\theoremstyle{definition}
\newtheorem{definition}[theorem]{Definition}

\theoremstyle{remark}

\let\d\partial
\let\n\noindent
\let\la\lambda
\def\D{{\mathcal D}}
\def\J{{\mathcal J}}
\def\K{{\mathcal K}}
\def\H{{\mathcal H}}
\def\I{{\mathcal I}}

\let\la\lambda
\let\La\Lambda
\begin{document}

\title{Jack polynomials in superspace}

\author{Patrick Desrosiers\thanks{pdesrosi@phy.ulaval.ca} \cr
\emph{D\'epartement de Physique, de G\'enie Physique et d'Optique},\cr
Universit\'e Laval, \cr
Qu\'ebec, Canada, G1K 7P4.
\and
Luc Lapointe\thanks{lapointe@inst-mat.utalca.cl }\cr
\emph{Instituto de Matem\'atica y F\'{\i}sica},\cr
Universidad de Talca,\cr
Casilla 747, Talca, Chile.
\and
Pierre Mathieu\thanks{pmathieu@phy.ulaval.ca} \cr
\emph{D\'epartement de Physique, de G\'enie Physique et d'Optique},\cr
Universit\'e Laval, \cr
Qu\'ebec, Canada, G1K 7P4.
}

\date{October 2002}

\maketitle

\begin{abstract}

This work initiates the study of {\it orthogonal} symmetric polynomials in
superspace. Here we present two approaches leading to
a family of orthogonal polynomials in superspace that generalize the
Jack  polynomials. The first approach relies on previous work by the
authors in which eigenfunctions of the supersymmetric extension of the
trigonometric Calogero-Moser-Sutherland Hamiltonian were constructed.
Orthogonal  eigenfunctions are now obtained by diagonalizing the first
nontrivial element of a bosonic tower of commuting conserved charges
not containing this Hamiltonian. Quite
remarkably, the
expansion coefficients of these orthogonal eigenfunctions in the
supermonomial basis are
stable with respect to the number of variables.
The second and more direct approach
amounts to symmetrize products of non-symmetric
Jack polynomials with monomials in the fermionic variables.  This
time, the orthogonality is inherited from the
orthogonality of  the non-symmetric Jack
polynomials, and the value of the norm is given explicitly.

\end{abstract}

\newpage

\section{Introduction}

A natural direction in which the theory of orthogonal
symmetric polynomials can be generalized is to consider its extension to
superspace.  One possible approach for such an extension is
to consider polynomials involving fermionic (i.e., Grassmannian) variables, or
superpolynomials,  that arise
from physically relevant eigenvalue problems invariant under
supersymmetry.

In many respects (physical and mathematical), one of the most
fundamental bases of symmetric  orthogonal polynomials is that of the Jack
polynomials. This work is concerned with their
orthogonality-preserving extension to superspace.

A basic requirement of Jack
superpolynomials is that they reduce to Jack
polynomials  when the
fermionic variables are set to zero.  Another requirement is that they
be solutions of
the supersymmetric
generalization of the eigenvalue problem characterizing  the Jack
polynomials.  More precisely, Jack
polynomials are eigenfunctions of the trigonometric
Calogero-Moser-Sutherland (tCMS) model (see e.g., \cite{LV} and
\cite{Stan,Mac} for properties of the Jack
polynomials). Jack
superpolynomials must thus be eigenfunctions of the supersymmetric
extension of the trigonometric Calogero-Moser-Sutherland (stCMS)
model \cite{DLM1}.\footnote{An extensive list of references on the CMS
model and its supersymmetric extension can be found in \cite{DLM1}.}

This eigenfunction characterization, as in the non-fermionic case,
does not uniquely
define Jack
superpolynomials. A triangular decomposition, in terms of a
superspace extension of the symmetric
monomial functions (or supermonomials), must be imposed.
Unique eigenfunctions, $\J_{\La}$, defined according to such a
triangular decomposition, were constructed in
\cite{DLM1,DLM2}, and called Jack superpolynomials.
They are
indexed by superpartitions $\La= (\La^a;\La^s)$, composed of a
partition $\La^a$ with
distinct parts, and a usual partition $\La^s$. The number of entries in $\La^a$
characterizes the fermion sector (i.e., the number of anticommuting
variables appearing in
every term of the expansion of $\J_{\La}$).

The integrability of
the  tCMS model also makes Jack polynomials eigenfunctions of a
family of $N$ independent commuting quantities, where $N$ is the number of
variables.  We prove in this article that, similarly,
Jack superpolynomials are eigenfunctions of a whole tower of
commuting conserved charges, denoted $H_n$, for $n=1,\cdots, N$,
where $H_2$ is the Hamiltonian of the stCSM model.  The proof relies heavily on
the remarkable fact that,
if we consider the restriction to the space of superpolynomials symmetric
under the simultaneous  interchange of any pair of
bosonic and fermionic variables, these
charges can be expressed
using Dunkl operators as $\H_n= \sum_{i=1}^N
(\D_i)^n$. That is, under this restriction,  $H_n$ is equivalent to $\H_n$.

Now, even though Jack superpolynomials are eigenfunctions of $N$
commuting conserved charges, degeneracies are still present.
Indeed,  two distinct superpolynomials labeled by two different
superpartitions built out of the same set of $N$ integers (but distributed
differently among the
two partitions $\La^a$ and $\La^s$)
have identical
$\H_n$ eigenvalues.
As a result, the Jack superpolynomials $\J_{\La}$ of \cite{DLM1,DLM2}
are {\it not orthogonal} under the scalar product \eqref{scalarp} with
respect to which
$\H_n$ is self-adjoint.

The Gram-Schmidt orthogonalization procedure can of course be used
to construct orthogonal superpolynomials, but a general pattern
is not likely to appear using this construction.  The question is thus
whether we can naturally define a family of
orthogonal superpolynomials.
The answer to this
question lies in the following observation. By extending the usual
tCMS model with
$N$ bosonic degrees of freedom to the supersymmetric case, we have
introduced $N$ new
degrees of freedom.  Integrability leads in this case to the appearance of
new conserved
charges. Indeed, there are $3N$ new conserved charges, new in the
sense that they all
disappear when the fermionic variables vanish
\cite{DLM1}. Among these new charges,
$2N$ of them are fermionic, that is, they change the fermion number of the
function on which they act.
However, the remaining
$N$ charges are bosonic, mutually commute, and do not
affect the fermion
number. These charges, denoted
$I_n$, $n=1,\dots,N,$ are thus natural candidates for extra
operators that may lift the degeneracy of the Jack superpolynomials,
and thereby produce orthogonal combinations
of these superpolynomials.

This expectation indeed materializes. Actually, to construct orthogonal
superpolynomials it suffices to consider the action of the charge
$I_1$, or  equivalently, its Dunkl-operator version $\I_1$ in
the space of symmetric superpolynomials.  Its action is also
triangular, but with respect
to an ordering on superpartitions stronger than the one introduced in
\cite{DLM2}. Knowing the  action of $\I_1$ explicitly on Jack superpolynomials
allows to define orthogonal fermionic
extensions, $J_{\La}$, of the usual Jack polynomials.
Moreover, it also leads to determinantal formulas for
     the expansion coefficients of the orthogonal Jack
superpolynomials $J_{\La}$ in terms of Jack superpolynomials $\J_\Omega$.

The program that we just sketched is the subject of the first
part of this paper (up to section 8).
% It constitutes one of the two
% constructions of the orthogonal Jack
% superpolynomials that we present here.
It is in line with our previous work \cite{DLM1,DLM2}, and can be viewed
as its natural completion.  We stress that it is also very explicit in that
the precise relation between the orthogonal Jack
superpolynomials $J_{\La}$  with the old $\J_{\La}$ is provided, and
that closed
form expressions for the latter, in the supermonomial basis, were already
obtained in \cite{DLM2}. The construction is also
`physical':  the quantum many-body problem and its
underlying integrability structure is the guiding tool used to
identify a complete set
of simultaneously diagonalizable operators.
%(give the plan of this part).

This leads us to our first characterization of the orthogonal Jack
superpolynomials:

\begin{theorem} (See Theorems~\ref{theo1} and \ref{theo6}).  The orthogonal
Jack
superpolynomial $J_{\La}$  is the  unique function satisfying:
\begin{equation}
\H_2 \, J_{\La} = \varepsilon_{\La} J_{\La} \, , \qquad
\I_1 \, J_{\La} = \epsilon_{\La} J_{\La} \, , \qquad {\rm and} \qquad
J_{\La} = m_{\La} + \sum_{\Omega < \Lambda} c_{\La \Omega}(\beta)\,
m_{\Omega} \, ,
\end{equation}
where $\varepsilon_{\La}$and $\epsilon_{\La}$ are defined in
Lemma~\ref{lemtril} and Theorem~\ref{bigtheo} respectively, while the
ordering on
superpartitions  is introduced in Definition~\ref{bruhatP}.
\end{theorem}

In the second part of the paper (which is essentially section 9) we propose a
much more direct, although less explicit, construction of the orthogonal Jack
superpolynomials. The starting point is not anymore the extension of the usual
Jack polynomial eigenvalue problem,  but rather a symmetrization process
performed on the non-symmetric Jack polynomials
\cite{Opdam,sahi}, suitably dressed with products of
fermionic variables.

Recall that the non-symmetric Jack polynomials $E_\la$ (where $\la$
is now a composition)
are eigenfunctions of the
$\D_i$ operators \cite{Opdam}, and, from the self-adjointness of these
operators, orthogonal. On the other
hand, as already
pointed  out, the stCMS commuting conserved charges can all be expressed in
term of the
$\D_i$'s.  This naturally suggests a very direct path for
the construction of the  common eigenfunctions of all the commuting stCMS
charges: start with
$E_\la$, add a fermionic-monomial prefactor and symmetrize with
respect to both types of
variables.  Quite remarkably, this indeed produces the orthogonal Jack
superpolynomials $J_{\La}$ (with
$\La$ determined by
$\la$ and the fermionic number).
The advantage of this construction is that orthogonality is built in, and
preserved by the symmetrization.
%  Notice that although this appraoch is very
% straithforward, it does not lead to closed form expressions comparable to the
% determinantal formulae which underly the first construction.

This second construction (especially the argument in the proof
of Theorem~\ref{jjbar}) leads to another characterization of
the orthogonal Jack superpolynomials:

\begin{theorem} (See Theorems~\ref{theoortho} and~\ref{jjbar}). The orthogonal
Jack superpolynomials are the unique functions satisfying:
\begin{equation}
\langle J_{\La}, J_{\Omega} \rangle_{\beta} \propto
\delta_{\Lambda \Omega} \, , \qquad {\rm and}
     \qquad J_{\La} = m_{\La} + \sum_{\Omega < \Lambda} c_{\La \Omega}(\beta)\,
m_{\Omega} \, .
\end{equation}
\end{theorem}

\n Our two characterizations of the orthogonal Jack superpolynomials
     thus extend
     the two most common definitions of Jack polynomials.

An important clarification is in order concerning the second
construction. Because of
the anticommuting nature of the fermionic variables, a symmetrized
superpolynomial that
contains
$m$ fermionic variables is necessarily an antisymmetric function of the $m$
corresponding bosonic variables, in addition to be symmetric in the
remaining bosonic
variables. In other words, viewed solely as functions of the bosonic
variables, the
symmetric superpolynomials can be decomposed as a sum over  polynomials with
mixed symmetry
properties such as those studied in \cite{Baker, Baker2}, but with coefficients
involving fermionic variables.
This, however, does
not mean that the theory of symmetric superpolynomials is only a
special case of the theory
of polynomials with mixed symmetry.  Indeed, by symmetrizing over all
the variables, including the fermionic ones, the antisymmetrized bosonic
variables will be dependent of the particular term under consideration,
ensuring that the net result is
a brand new object.
Let us illustrate this comment by comparing  a four-variable  monomial
function with mixed symmetry with its supermonomial counterpart.  Take the
four variables to be $x_1, x_2$, and $y_1,y_2$, and let the $x_i$ variables  be
symmetrized while the $y_i$ are antisymmetrized. Let also the $x$ and $y$
parts of the monomial be parametrized by the partitions $(2,1)$ and $(3,1)$
respectively. The  associated monomial with mixed symmetry reads:
\begin{equation}\label{exmono}
{\widetilde m}_{(3,1)_a,(2,1)_s}=(y_1^3y_2-y_1y_2^3)(x_1^2x_2+x_1x_2^2)
\end{equation}
The corresponding supermonomial
has four bosonic and four
fermionic  variables denoted $z_i$ and $\theta_i$ respectively, with
$i=1,\cdots, 4$ and $\theta_i\theta_j=-\theta_j\theta_i$. Given
the superpartition $(3,1;2,1)$, it reads as
\begin{eqnarray}\label{exmo}
m_{(3,1;2,1)}=~&\theta_1\theta_2(z_1^3z_2-z_1z_2^3)(z_3^2z_4+z_3z_4^2)+
\theta_1\theta_3(z_1^3z_3-z_1z_3^3)(z_2^2z_4+z_2z_4^2)  \nonumber \\
+ &\theta_1\theta_4(z_1^3z_4-z_1z_4^3)(z_2^2z_3+z_2z_3^2)+
\theta_2\theta_3(z_2^3z_3-z_2z_3^3)(z_1^2z_4+z_1z_4^2)\\ \nonumber
+ &\theta_2\theta_4(z_2^3z_4-z_2z_4^3)(z_1^2z_3+z_1z_3^2)+
\theta_3\theta_4(z_3^3z_4-z_3z_4^3)(z_1^2z_2+z_1z_2^2)
\end{eqnarray}
Clearly, each bosonic component of this expression corresponds to a
monomial with mixed symmetry of type (\ref{exmono}). The main
point is that the supermonomial is the sum of all these mixed symmetry
monomials, each with its appropriate fermionic-monomial prefactor.  These
prefactors have  drastic effects say, when multiplying  supermonomials
together, due to the fermionic nature of their constituents.  The
multiplication of
polynomials reveals in a rather critical way one aspect of the
difference between the
     polynomials with mixed symmetry and superpolynomials: the product of two
polynomials of the former type cannot be decomposed into a linear
combination of
polynomials with mixed symmetry, that is, there is no ring structure.
This is not so for
the superpolynomials.
It should thus be crystal clear that Jack superpolynomials are
not simply Jack polynomials with mixed symmetry properties in
disguised  form. However,
pinpointing the relationship between these two types of objects  is
technically important
since it allows to use the results of
\cite{Baker} on the norm of the Jack polynomials with mixed symmetry to
obtain, in a rather direct way, the norm of the  orthogonal
Jack superpolynomials.

The presentation of the two approaches is preceded by four sections
in which we introduce our notation, define our basic superobjects and derive
relevant properties. Section 6 and 7 deal respectively with
the construction of the
$\H_n$ and $\I_n$ eigenfunctions.  The common eigenfunctions are shown to be
orthogonal in Section 8. The alternative construction based on the
non-symmetric Jack polynomials is the subject of Section 9.

In Appendix A, we present a number
of examples of Jack
superpolynomials, including a detailed
computation based on the
determinantal formula. These examples illustrate a nice property of the
orthogonal Jack
superpolynomials:  they do not depend upon $N$ (when $N$
is sufficiently
large). In other words, their expansion coefficients in the
supermonomial basis are
independent of the number of variables. This property can of course be
obtained from the explicit
formulas, but it is
not at once manifest.

  Finally, various natural extensions of this work are
mentioned in the
conclusion.

Note that for the readers not particularly interested in
the `physical' construction  relying on the structure of the integrable
supersymmetric stCMS model and its conserved charges,
reading Section 2, and the first two definitions of Section 4 is
sufficient to understand Section 9.

\section{Basic definitions}

For $i,j \in \{1,\dots,N \}$,
let $K_{ij}$
be the operator that exchanges the variables $z_i$ and $z_j$.  Similarly,
let $\kappa_{ij}$ exchange the anticommuting variables
$\theta_i$ and $\theta_j$.  Their action on a superfunction
$f(z,\theta)$ is thus
\begin{eqnarray}
& \, \, K_{ij}f(\dots, z_i, \dots, z_j, \dots, \theta_i, \dots
\theta_j, \dots)= f(\dots, z_j,
\dots, z_i,
\dots, \theta_i,\dots, \theta_j,\dots)\; ,\nonumber\\
& \kappa_{ij}
f(\dots, z_i, \dots, z_j, \dots, \theta_i,\dots,
\theta_j,\dots)= f(\dots,
z_i,
\dots, z_j,
\dots, \theta_j, \dots, \theta_i,\dots)\;.
\end{eqnarray}

Each
of these sets of operators generates a realization of the permutation
group $S_N$.
Since the ${K}_{ij}$'s and the $\kappa_{ij}$'s commute,
the operators ${\mathcal K}_{ij}= K_{ij} \kappa_{ij}$, acting as
\begin{equation}
{\mathcal K}_{ij}f(\dots, z_i, \dots, z_j, \dots, \theta_i,\dots, \theta_j,
\dots)=
f(\dots, z_j,
\dots, z_i,
\dots, \theta_j,\dots, \theta_i,\dots)\;,
\end{equation}
      are also seen
to generate a realization
of $S_N$.

Let $P$ be the space of polynomials in the variables $\theta_1,\dots,\theta_N$
and $z_1,\dots,z_N$.
We will denote by $P^{S_N}$ the subspace of $P$
of polynomials invariant under the simultaneous
exchange of any pair of variables $\theta_i \leftrightarrow \theta_j$ and
$z_i \leftrightarrow z_j$.  A polynomial $f\in P$ thus belongs
to $P^{S_N}$ iff ${\mathcal K}_{\sigma} \, f = f$ for any $\sigma \in
S_N$.\footnote{It is understood that the product decomposition of
$\K_\sigma$ into
elementary permutations $\K_{i,i+1}$ follows the decomposition of $\sigma$ into
elementary transpositions $\sigma_i=(i,i+1)$.  In other words, $\K_\sigma$ is a
realization on superspace variables of the action of $\sigma$ on indices.}

Let $I=\{i_1,\dots,i_m \}$, ($1\leq i_1 < i_2 < \cdots < i_m \leq N$) be an
ordered set of integers, and let $\lambda$ be a
composition with $N$ parts,
that is, a sequence of
$N$ nonnegative integers ({\it e.g.}, if $N=5$, one possible composition is
(20134))\footnote{We use the word composition in a broader sense, given that,
strictly speaking, a composition should not contain any zeroes.  Similarly,
we allow the presence of zeroes in a partition.}.
A natural basis
of $P$ is provided by the
monomials $\theta_I \, z^{\lambda}$, where
\begin{equation}
\theta_I=\theta_{\{i_1,\dots,i_m \}}= \theta_{i_1}\cdots \theta_{i_m}\; ,
\qquad z^{\lambda}= z_1^{\la_1}\cdots z_N^{\la_N} \,.
\end{equation}
If $I$
has
$m$ entries, $\theta_I z^{\lambda}$
is said to belong to the $m$-fermion sector.

A superpartition $\La$ in the $m$-fermion sector is made of a partition
$\La^a$ whose parts are all distinct, and of a usual partition
$\La^s$, that is,
\begin{equation}
\Lambda=(\Lambda_1,\ldots,\Lambda_m;\Lambda_{m+1},\ldots,\Lambda_{N})
=(\La^a;\La^s) \, ,
\end{equation}
with
\begin{equation}
\La^a=(\Lambda_1,\ldots,\Lambda_m), \qquad
\Lambda_i>\Lambda_{i+1}\geq 0 \,,\quad \  i=1, \ldots m-1,
\end{equation}
and
\begin{equation}
\La^s= (\Lambda_{m+1},\ldots,\Lambda_{N}),  \qquad
\Lambda_i \ge \Lambda_{i+1}\geq 0 \, , \quad  i=m+1,\dots,N-1 \, .
% \Lambda_i \geq 0\mbox{ if }i=m+1\mbox{ ; }
% \Lambda_i > 0  \quad\forall  i> m+1.
\end{equation}
In the zero-fermion sector, the
semicolon is omitted and
         $ \Lambda $  reduces to  $ \La^s $. We often write the degree of a
superpartition as  $ n =|\Lambda|=\sum_{i=1}^{N}\Lambda_i$.
Given a supercomposition $\gamma=(\gamma_a;\gamma_s)=(\gamma_1,\dots,\gamma_m;
\gamma_{m+1},\dots,\gamma_N)$, we will denote by $\overline{\gamma}$
the superpartition whose antisymmetric part is the
rearrangement of $(\gamma_1,\dots,\gamma_m)$ and whose symmetric part is
the rearrangement of $(\gamma_{m+1},\dots,\gamma_N)$. Denoting by $\lambda^+$
the partition obtained by the rearrangement of the entries of any composition
$\lambda$, we have
\begin{equation}
\overline{(\gamma_a;\gamma_s)} = (\gamma_a^+;\gamma_s^+)\, ,
\end{equation}
which we can illustrate with the example:
\begin{equation}
\overline{(1,4,2;2,5,1,3)}
=(4,2,1;5,3,2,1).
\end{equation}
Furthermore, $\sigma_{\gamma}$ will stand for the element
of $S_N$
that sends $\gamma$ to $\overline{\gamma}$, that is $\sigma_{\gamma} \gamma=
\overline{\gamma}$.  Note that we can always choose
$\sigma_{\gamma}$ such that
$\sigma_{\gamma}=\sigma_{\gamma}^a\sigma_{\gamma}^s$,
with $\sigma_{\gamma}^a$ and $\sigma_{\gamma}^s$ permutations of
$\{1,\dots,m\}$
and $\{m+1,\dots,N \}$ respectively.

If we delete the semi-colon in a superpartition $\La$, we obtain an ordinary
composition that we will denote as $\La_c$:
\begin{equation}
\Lambda=(\Lambda_1,\ldots,\Lambda_m;\Lambda_{m+1},\ldots,\Lambda_{N})
\qquad \Longrightarrow\qquad\Lambda_c=(\Lambda_1,\ldots,\Lambda_{N}) \, .
\end{equation}
Finally, to any  superpartition  $ \Lambda $,  we associate a unique
standard partition  $ \Lambda^* $  obtained by rearranging the parts of the
superpartition in decreasing order:
\begin{equation}
\Lambda^*=(\Lambda_c)^+\;.
\end{equation} For instance, the $*$-rearrangement of
        $(4,2,1;5,3,2,1)$ is
\begin{equation}
(4,2,1;5,3,2,1)^*
=(5,4,3,2,2,1,1) \, .
\end{equation}

A natural basis of $P^{S_N}$ is provided by the monomial symmetric
superfunctions\footnote{The quantity $z^{\Lambda}$ is to be understood as
$z_1^{\Lambda_1}\cdots z_N^{\Lambda_N}$, that is, as if $\La$ were replaced by
$\La_c$. However, to alleviate the notation,  we will omit the
subindex $c$ when $\La$ is treated as a formal power.}
\begin{equation} \label{mono}
m_{\Lambda}= \frac{1}{f_{\Lambda}} \sum_{\sigma \in S_N} {\mathcal
K}_{\sigma}
\, \theta_{\{1,\dots,m \}} z^{\Lambda} \, ,
\end{equation}
where the normalization constant $f_{\Lambda}$ is
\begin{equation}
f_{\Lambda} = f_{\La^s}= n_{\La^s}(0)!\, n_{\La^s}(1)! \,
n_{\La^s}(2) ! \cdots \, ,
\end{equation}
with  $n_{\La^s}(i)$ the number of $i$'s in $\La^s$, the
symmetric part of
the superpartition
$\Lambda=(\La^a;\La^s)$.  This normalization ensures
that the coefficient of the
monomial $\theta_{\{1,\dots,m \}} \, z^{\Lambda}$ appearing
in the expansion of $m_{\Lambda}$ is equal to $1$. The supermonomial
$m_{(3,1;2,1)}$
is given in (\ref{exmo}) for $N=4$.

Finally, we will define a scalar product $\langle.,.\rangle_{\beta}$ in $P$.
With
\begin{equation}
\Delta(z) = \prod_{1\leq j<k \leq N} \left[ \frac{z_j-z_k}{z_j z_k}
\right] \, ,
\end{equation}
$\langle.,.\rangle_{\beta}$ is defined (for $\beta$ a positive integer)
on the basis elements of $P$ as
\begin{equation} \label{scalarp}
\langle \, \theta_I z^{\lambda},\theta_{J}z^{\mu} \, \rangle_{\beta} =
\begin{cases}
{\rm C.T.} \left[\Delta^{\beta}(z) \Delta^{\beta}(\bar z)  z^{\lambda}/
z^{\mu}
         \right] & {\rm if~} I=J \\
0 & {\rm otherwise}
\end{cases} \, ,
\end{equation}
where  ${\bar z}_i= 1/z_i$, and  where ${\rm C.T.}[E]$ stands for the
constant term of the expression $E$. This scalar product is a special
case of the
physical scalar product of the  underlying supersymmetric quantum many-body
problem ($\beta$ is now arbitrary)
\begin{equation}
\langle \, A(z,\theta), B(z,\theta) \, \rangle_\beta=
\left(\prod_{i=1}^N \oint \frac{d z_i}{2\pi
iz_i}\right) \int d\theta_1\cdots d\theta_N \Delta^{\beta}(z)
\Delta^{\beta}(\bar z) A(z,\theta), B({\bar z},{\bar \theta}) \, ,
\end{equation}
where $\overline{\theta_{i_1}\cdots \theta_{i_m}}$ is defined such that
\begin{equation}
(\theta_{i_1}\cdots \theta_{i_m})(\overline{\theta_{i_1}\cdots \theta_{i_m}})=
\theta_{N}\cdots \theta_{1} \, ,
\end{equation}
an operation akin to the Hodge duality transformation. For instance,
if $N=5$, we have
$\overline{\theta_{2} \theta_{5}}=
-\theta_{4} \theta_{3} \theta_{1}
$. The integral over fermionic variables refers to the Berezin
integration
\begin{equation}
\int d\theta= 0\; , \qquad \int d\theta \, \theta = 1\, .
\end{equation}

\section{Dunkl operators and conserved charges}

%{\bf Dunkl, Dunkl-Cherednik, Opdam-Cherednik???}

The Dunkl operators, ${\mathcal D}_i$, are defined as
\cite{Chered}\footnote{Following \cite{DLM1}, we use the qualitative
`Dunkl' for all
Dunkl-type operators. The present $\D_i$'s are often
called
Cherednik operators.}
\begin{eqnarray} \label{defDi}
{\mathcal D}_i &=& z_i \frac{\partial}{\partial z_i} +
\beta \sum_{j< i} \frac{z_i}{z_i-z_j} \left( 1-K_{ij} \right) +
\beta \sum_{j> i} \frac{z_j}{z_i-z_j} \left( 1-K_{ij} \right) -\beta (i-1)
\nonumber \\
&=& z_i \frac{\partial}{\partial z_i} +
\beta \sum_{j<i} {\mathcal  O}_{ij}+
\beta \sum_{j> i} {\mathcal  O}_{ij} -\beta (i-1)
\, ,
\end{eqnarray}
where
\begin{equation}
{\mathcal O}_{ij} = \begin{cases}
\frac{z_i}{z_i-z_j} \left( 1-K_{ij} \right) & j<i \\
\frac{z_j}{z_i-z_j} \left( 1-K_{ij} \right) & j> i
\end{cases} \, .
\end{equation}
The set $\{{\mathcal D}_i\}$ forms a family of commuting operators
satisfying the degenerate Hecke relations
\begin{eqnarray} \label{hecke}
K_{i,i+1} {\mathcal D}_{i+1} - {\mathcal D}_i K_{i,i+1} = \beta
\qquad {\rm and} \qquad K_{j,j+1} {\mathcal D}_{i} = {\mathcal D}_i
K_{j,j+1} \quad
        (i \neq j,j+1) \, .
\end{eqnarray}
It turns out that any conserved charge, $C_n$, of the stCMS model can be
written as the $P^{S_N}$-projection of an expression, ${\mathcal C}_n$,
involving Dunkl operators
\begin{equation}
{\mathcal C}_n|_{P^{S_N}} = C_n \qquad \Longleftrightarrow\qquad
{\mathcal C}_n\, f = C_n \, f \; , \qquad \forall f\in P^{S_N}\, .
\end{equation}
This is a key tool in our subsequent analysis,
since by working with Dunkl operators, we
avoid manipulating fermionic variables to a large extent.
More explicitly, the stCMS conserved charges are defined  as  the
projection onto ${P^{S_N}}$ of the following expressions
%\begin{equation}
\begin{eqnarray}
{\mathcal H}_n &= &\sum_{i=1}^N
        {\mathcal D}_i^n  \, ,\\
{\mathcal Q}_n &= &\sum_{w \in S_N} {\mathcal K}_{w}
\left( \theta_1  {\mathcal D}_1^n  \right)  \, ,\\
{\mathcal Q}_n^\dagger &= &\sum_{w \in S_N} {\mathcal K}_{w}
\left(  \frac{\d}{ \d \theta_1} {\mathcal D}_1^n  \right)  \,
,\\ {\mathcal I}_n &= &\sum_{w \in S_N} {\mathcal K}_{w}
\left( \theta_1 \frac{\d}{ \d \theta_1} {\mathcal D}_1^n
\right)  \, .
\end{eqnarray}
%\end{equation}
% \left|_{P^{S_N}}\right.
Of these charges, $\I_0/(N-1)!$ gives the fermion number.
Observe that ${\mathcal
H}_n $ and $
{\mathcal I}_n $ preserve the number of fermions (the number of
$\theta_i$'s) of the superpolynomials on which they act,
while ${\mathcal Q}_n$ (resp. ${\mathcal Q}_n^\dagger$) increases
it by $1$ (resp. $-1$).  Also, since ${\mathcal H}_n$ is known to
be central in the degenerate Hecke algebra \cite{Opdam},
it commutes with $K_{\sigma}$,
for any
element $\sigma$ of the symmetric group.
 Finally,
these expressions not being unique,
we should mention that the present choice for ${\mathcal
H}_n $ and ${\mathcal I}_n $ is motivated by the requirement
that they act triangularly on monomial superfunctions (as we will
show later on).

We finish this section with two propositions, the first one relying on
the following lemma.
\begin{lemma} \label {petitlemme} For any nonnegative integers $n$ and $m$,
we have
\begin{equation}
[\D_1^n \K_{12},\D_1^m \K_{12}]+
[\K_{12} \D_1^n ,\K_{12} \D_1^m]=0 \, .
\end{equation}
\end{lemma}
\n {\it Proof}.  Given that $\kappa_{12}$ (in $\K_{12}$)
commutes with $\D_1$, the lemma is equivalent to
\begin{equation} \label{aprouver2}
[\D_1^n K_{12},\D_1^m K_{12}]+
[K_{12} \D_1^n ,K_{12} \D_1^m]=0 \, .
\end{equation}
We will now seek to prove this expression.  First, it is easy
to verify, using $K_{12}{\mathcal D}_1 ={\mathcal D}_2 K_{12} - \beta$,
that
\begin{equation} \label{idenDK}
K_{12} {\mathcal D}_1^n ={\mathcal D}_2^n K_{12} - \beta (
{\mathcal D}_1^{n-1}+{\mathcal D}_1^{n-2} {\mathcal D}_2+ \cdots
{\mathcal D}_1 {\mathcal D}_2^{n-2}+{\mathcal D}_2^{n-1})=
{\mathcal D}_2^n K_{12} -\beta h_{n-1}({\mathcal D}_1,{\mathcal D}_2) \, ,
\end{equation}
where $h_i(x_1,x_2)$ is the homogenous symmetric function of degree $i$ in
the variables $x_1$ and $x_2$.

Using \eqref{idenDK} and the commutativity of
${\mathcal D}_1$ and ${\mathcal D}_2$, we obtain
\begin{eqnarray}
[\D_1^n K_{12},\D_1^m K_{12}]+
[K_{12} \D_1^n ,K_{12} \D_1^m] & = & {\mathcal D}_1^n {\mathcal D}_2^m - \beta h_{m-1}  {\mathcal D}_1^n K_{12} -
{\mathcal D}_1^m {\mathcal D}_2^n + \beta h_{n-1} {\mathcal D}_1^m K_{12}
 \nonumber\\
& ~ & \qquad +
 {\mathcal D}_2^n {\mathcal D}_1^m - \beta h_{n-1} K_{12} {\mathcal D}_1^m - {\mathcal D}_2^m {\mathcal D}_1^n + \beta h_{m-1} K_{12} {\mathcal D}_1^n
\nonumber\\
   & = & - \beta h_{m-1}  {\mathcal D}_1^n K_{12} +
\beta h_{n-1}  {\mathcal D}_1^m K_{12} -
\beta h_{n-1}  {\mathcal D}_2^m K_{12}
+ \beta^2 h_{n-1} h_{m-1}
 \nonumber\\
& ~ & \qquad +
 \beta h_{m-1}  {\mathcal D}_2^n K_{12}
- \beta^2 h_{n-1} h_{m-1} \nonumber \\
&=& \beta h_{m-1} ({\mathcal D}_2^n - {\mathcal D}_1^n) K_{12} +
 \beta h_{n-1} ({\mathcal D}_1^m - {\mathcal D}_2^m) K_{12} \, ,
\end{eqnarray}
where $h_i$ stands for $h_i({\mathcal D}_1,{\mathcal D}_2)$.
Finally, using the simple identity
$
\left({x}_1^m
-{x}_2^m \right) = \left({x}_1
-{x}_2 \right) h_{m-1}({x}_1,{x}_2)
$, the previous expresion vanishes, and \eqref{aprouver2} is thus
seen to hold.
\hfill $\square$

\begin{proposition} The families of operators
${\mathcal H}_n$, $n=1,\dots,N$, and ${\mathcal I}_m$, $m=1,\dots,N$,
when acting on $P^{S_N}$,
form a set of mutually commuting operators, that is, they satisfy
\begin{equation}
[\H_n,\H_m] f=[\H_n,\I_m] f =[\I_n,\I_m]  f= 0 \, ,
\end{equation}
for any $f \in P^{S_N}$.
\end{proposition}

\n {\it Proof}. Since the Dunkl operators $\D_i$ mutually commute, we have
immediately $[\H_n,\H_m]=0$.  Further, since $\H_n$ commutes with $\K_{\sigma}$
for any permutation $\sigma$, we also get $[\H_n,\I_m]=0$.
The relation
$[\I_n,\I_m]f= 0$ is less trivial. We have
\begin{eqnarray}
\I_n \I_m f &=& \sum_{w,\sigma \in S_N} \K_{w} \, \theta_1
\frac{\d}{ \d \theta_1} \D_1^n \, \K_{\sigma} \, \theta_1
\frac{\d}{ \d \theta_1} \D_1^m \, f  \nonumber \\
        &=& \sum_{w,\sigma \in S_N} \theta_{(w)_1}
\frac{\d}{ \d \theta_{(w)_1}}  \theta_{(w\sigma)_1}
\frac{\d}{ \d \theta_{(w\sigma)_1}}
\K_{w} \,  \D_1^n \, \K_{\sigma} \,
\D_1^m \, f \, ,
\end{eqnarray}
where $(w)_1$ is the first entry of the permutation $w$.
Therefore, $[\I_n,\I_m]f$ can be written as
\begin{equation}
[\I_n,\I_m]f = \sum_{w,\sigma \in S_N} \theta_{(w)_1}
\frac{\d}{ \d \theta_{(w)_1}}  \theta_{(w\sigma)_1}
\frac{\d}{ \d \theta_{(w\sigma)_1}} \Bigr(
\K_{w} \,  \D_1^n \, \K_{\sigma} \,
\D_1^m  - \K_{w} \,  \D_1^m \, \K_{\sigma} \,
\D_1^n \Bigl) \, f \, .
\label{commuf}
\end{equation}
To prove that this expression is equal to zero, we will match its
summands $(w,\sigma)$ and  $(w\sigma, \sigma^{-1})$, and see that
they cancel each others.  First, if we let $w \to w \sigma$, and $\sigma \to
\sigma^{-1}$, the summand $(w,\sigma)$ of \eqref{commuf} becomes
\begin{equation}
\begin{split}
& \theta_{(w\sigma)_1}\frac{\d}{ \d \theta_{(w\sigma)_1}}
        \theta_{(w)_1}
\frac{\d}{ \d \theta_{(w)_1}} \Bigr(
\K_{w} \K_{\sigma} \,  \D_1^n \, \K_{\sigma^{-1}} \,
\D_1^m  - \K_{w} \K_{\sigma} \,  \D_1^m \, \K_{\sigma^{-1}} \,
\D_1^n \Bigl) \, f  \\
& \qquad \qquad = \theta_{(w)_1}\frac{\d}{ \d \theta_{(w)_1}}
        \theta_{(w \sigma)_1}
\frac{\d}{ \d \theta_{(w \sigma)_1}} \Bigr(
\K_{w} \K_{\sigma} \,  \D_1^n \, \K_{\sigma^{-1}} \,
\D_1^m  - \K_{w} \K_{\sigma} \,  \D_1^m \, \K_{\sigma^{-1}} \,
\D_1^n \Bigl) \, f \, ,
\end{split}
\label{autre}
\end{equation}
the equality being obtained by interchanging the two prefactors
$\theta_i\d/\d\theta_i$.
Now,
$(w,\sigma)=(w
\sigma,\sigma^{-1})$ iff
$\sigma=e$. Since, in the case $(w,\sigma)=(w,e)$,
the summand
$(w,\sigma)$ of \eqref{commuf} cancels, we can assume that
$(w,\sigma)$ and $(w \sigma,\sigma^{-1})$ are distinct summands.
Having shown that the prefactors are the same for the two summands
$(w,\sigma)$ and
$(w\sigma, \sigma^{-1})$ (cf.
\eqref{autre}), verifying their cancellation amounts to checking that
\begin{equation}
\Bigr( \K_{w} \,  \D_1^n \, \K_{\sigma} \,
\D_1^m  - \K_{w} \,  \D_1^m \, \K_{\sigma} \,
\D_1^n \Bigl) \, f \, + \,  \Bigr(
\K_{w} \K_{\sigma} \,  \D_1^n \, \K_{\sigma^{-1}} \,
\D_1^m  - \K_{w} \K_{\sigma} \,  \D_1^m \, \K_{\sigma^{-1}} \,
\D_1^n \Bigl) \, f = 0 \, .
\end{equation}
Since $f \in P^{S_N}$, this is equivalent to
\begin{equation}
\Bigr(  \D_1^n \, \K_{\sigma} \,
\D_1^m \K_{\sigma^{-1}} -   \D_1^m \, \K_{\sigma} \,
\D_1^n \K_{\sigma^{-1}} +
       \K_{\sigma} \,  \D_1^n \, \K_{\sigma^{-1}} \,
\D_1^m  -  \K_{\sigma} \,  \D_1^m \, \K_{\sigma^{-1}} \,
\D_1^n \Bigl) \, f = 0 \, .
\label{aprouver}
\end{equation}
Now, $\D_1$ commutes with $\K_{i,i+1}$, as
long as $i \neq 1$. Therefore, if $\sigma$ leaves $1$ fixed, \eqref{aprouver}
holds. We can thus assume that $\sigma$ does not leave $1$ fixed.  In this
case,
$\sigma$ can be decomposed as
$\alpha (12) \beta$, where $\alpha$ and $\beta$ are permutations that
leave $1$ invariant, and \eqref{aprouver} becomes
\begin{equation}
\K_{\alpha} \Bigr(  \D_1^n \, \K_{12} \,
\D_1^m \K_{12}  -    \D_1^m \, \K_{12} \,
\D_1^n \K_{12}
+  \K_{12} \,  \D_1^n \, \K_{12} \,
\D_1^m   -   \K_{12} \,  \D_1^m \, \K_{12} \,
\D_1^n
\Bigl) \K_{\alpha^{-1}} \, f = 0 \, ,
\label{aprouverfinal}
\end{equation}
where we have used the facts that $\K_{12}^{-1}=\K_{12}$, and that
$\K_{\alpha}$ and $\K_{\beta}$ commute with $\D_1$.
Given that from Lemma~\ref{petitlemme},
$[\D_1^n \K_{12},\D_1^m \K_{12}]+
[\K_{12} \D_1^n ,\K_{12} \D_1^m]=0$, the expression is finally
seen to hold,
thereby proving
$[\I_n,\I_m] f= 0$.
\hfill $\square$

\begin{proposition}
The charges $\H_n$ and $\I_n$ are  self-adjoint with
respect to the scalar product \eqref{scalarp}.
\end{proposition}

\n {\it Proof}. In the $\H_n$ case, this simply follows from
      the self-adjointness of the
operators $\D_i$ \cite{Opdam}.  In the case of $\I_n$, we also need to use
$\theta_1^\dagger=\frac{\d}{\d\theta_1}$, and
\begin{equation}
\left(\theta_1\frac{\d}{\d\theta_1}\right)^\dagger=
\left(\frac{\d}{\d\theta_1}\right)^\dagger
\theta_1^\dagger= \theta_1\frac{\d}{\d\theta_1} \, .
\end{equation}
\hfill $\square$

Our first goal will be to find the common eigenfunctions of the
commuting operators
${\mathcal H}_n $
and
$
{\mathcal I}_n $.  However, before plunging into the relevant
computations, we need to introduce further technical tools.  This will be
the subject of
the following
two sections.

\section{Orderings on superpartitions}

In this section we introduce three orderings on superpartitions.  They will
provide three different ways of defining triangular
decompositions.

First recall the usual dominance ordering on
partitions \cite{Mac}. If $\la$ and $\mu$ are two partitions (i.e., $\la=\la^+$
and
$\mu=\mu^+$), then
$\lambda
\geq
\mu$ iff
$\lambda_1+
\cdots +\lambda_i
\geq
\mu_1+ \cdots +\mu_i$ for all $i$. This ordering can be extended to
compositions as follows. Any composition $\la$ can be obtained from its
rearranged partition $\la^+$ by a sequence of permutations.
Among all permutations $w$ such that $\lambda= w \lambda^+$, there
exists a unique one, denoted $w_{\lambda}$, of minimal length.

\begin{definition} \label{bruhat}
Given two compositions $\la, \mu$, we say that $\lambda \geq \mu $
if either $\lambda^+ > \mu^+$, or $\lambda^+=\mu^+$ and
$w_{\lambda} \leq w_{\mu}$ in the Bruhat order of the symmetric
group. This will be called the {\it Bruhat ordering on
compositions}.\footnote{The ordering on compositions could
alternatively be formulated as follows \cite{Baker}. We say that
$\lambda \geq \mu $ if either $\lambda^+ > \mu^+$, or
$\lambda^+=\mu^+$ and $\sum_{i=1}^k\lambda_i\geq
\sum_{i=1}^k\mu_i$ for all $k$.}
\end{definition}

An immediate consequence of this definition is that $\lambda^+ \geq \mu$ for any
composition $\mu$ such that $\mu^+=\lambda^+$.
Moreover, given that to any
superpartition is associated  a composition, this ordering induces
an ordering on superpartitions.

\begin{definition} \label{bruhatP}
Given two superpartitions $\La, \Omega$, we say that
$\Lambda
\geq
\Omega$, if
$\Lambda_c \geq \Omega_c$. We shall refer to this ordering as
the {\it Bruhat ordering
on superpartitions}.
\end{definition}

We finally define two other orderings on superpartitions.

\begin{definition} \label{rorder}
The {\it $h$-ordering} $\leq_h$ is  defined such that $\Omega \leq_h \Lambda$,
if either
$\Lambda=\Omega$, or $\Omega^* \neq \Lambda^*$ and
$\Omega \leq \Lambda$.
\end{definition}

\begin{definition} \label{torder}
The {\it $t$-ordering} $\leq_t$
is  defined such that $\Omega \leq_t \Lambda$, iff $\Lambda^*=\Omega^*$ and
$\Omega \leq \Lambda$.
\end{definition}

Obviously, these two new orderings on superpartitions are special cases of the
Bruhat ordering
       in the sense that
       if either
$\Omega \leq_h \Lambda$ or $\Omega \leq_t \Lambda$, then
$\Omega \leq \Lambda$.

Let us look at illustrative examples. The two superpartitions
$\Omega=(5,3;4,1,1)$
and
$\La=(5,1;4,4,0)$ cannot be $t$-compared since $\Omega^*\not=\La^*$.
However, they can be
$h$-compared since
$(\Omega_c)^+=\Omega^* = (5,4,3,1,1)<(\La_c)^+=\La^*=(5,4,4,1,0)$.
On the other hand, let us see how $\Gamma=(5,1;4,3,1)$ compares with the
previous two superpartitions. Again, $\Gamma$ and $\La$ cannot be
$t$-compared, but
are such that $\Gamma<_h\La$. Since $\Gamma^*=\Omega^*$, the two
superpartitions
$\Gamma$ and $\Omega$
may be
$t$-comparable.  With
$\Gamma_c=(5,1,4,3,1)
=\sigma_2\sigma_3(5,4,3,1,1)$ and
$\Omega_c=(5,3,4,1,1)=\sigma_2(5,4,3,1,1)$, we in fact conclude that
$\Gamma<_t\Omega$.

As we will see later, the $h$-ordering characterizes the triangular  action
of the charges
$\H_n$ on the supermonomial basis, thus justifying its labeling
$h$.\footnote{This is the ordering introduced in \cite{DLM1}. In
reference \cite{DLM2}, a more precise formulation of this ordering was
introduced (and called
$\leq_s$). }  On the other hand, the action of the charges $\I_n$
on the non-orthogonal Jack superpolynomials
will be shown
to be triangular with respect to a different ordering. This
will happen to be the
$t$-ordering.
      The label $t$ refers in this case to the
interchange operator $T$, which we introduce below, relating
the superpartitions
that can be $t$-compared.

The rest of this section is devoted to presenting some properties of the Bruhat
ordering on compositions. Remarkably, the action of a Dunkl operator $\D_i$ is
triangular with respect to this ordering.
\begin{property} \label{proptriDi} \cite{Opdam, sahi}.
       Let $\lambda$ be a composition.  Then,
\begin{equation}
\D_i \, z^{\lambda} = \underline{\lambda_i} z^{\lambda} +
\sum_{\mu < \lambda} c_{\lambda \mu} \, z^{\mu} \, ,
\end{equation}
where
\begin{equation} \label{lsous}
\underline{{\lambda}_i}= \lambda_i-\beta
\Bigl( \# \{j=1,...,i-1 \, | \, \lambda_j \geq \lambda_i \}
+ \# \{j=i+1,...,N \, |\,  \lambda_j > \lambda_i \}
\Bigr) \, .
\end{equation}
\end{property}

For $i<j$, let $T_{ij}$ be such that on a composition $\lambda$,
\begin{equation}
T_{ij} \lambda =
\left \{
\begin{matrix}
(\cdots \lambda_j \cdots \lambda_i \cdots )
        & {\rm if~} \lambda_i > \lambda_j \\
(\cdots \lambda_i \cdots \lambda_j \cdots)
& {\rm otherwise}  \\
\end{matrix}
\right. \, ,
\end{equation}
that is, $T_{ij}$ interchanges the entries $\lambda_i$ and
$\lambda_j$ only when $\lambda_i > \lambda_j$.  The action of $T_{ij}$ on
superpartitions $\La$ is defined via the corresponding compositions $\La_c$.
The order on compositions satisfies the following obvious property.

\begin{property} Let $\mu$ and $\lambda$ be two compositions such that
       $\mu \leq \lambda$, and $\mu^+=\la^+$, that is, such that
$\mu\leq_t\lambda$.  Then,
there exists a sequence of operators $T_{ij}$ giving
\begin{equation} \label{applica}
\mu = T_{i_1 j_1} \cdots T_{i_\ell j_\ell} \lambda \, .
\end{equation}
\end{property}
\begin{lemma} \label{lempasimpo}
Let $\Omega$ and $\Lambda$ be two superpartitions.  If
\begin{equation} \label{eqorder}
\Omega = \overline{T_{i_1 j_1} \cdots T_{i_\ell j_\ell} \Lambda} \, ,
\end{equation}
for some $T_{i_k j_k}$, $1\leq k \leq \ell$, then $\Omega \leq \Lambda$.
\end{lemma}

First, it is important to realize that the product
$T_{i_1 j_1} \cdots T_{i_\ell j_\ell}$ can be rewritten as a product
of $T_{ij}$'s where all the $T_{ij}$'s that
interchange elements between the fermionic and bosonic sectors (that is, that
interchange entries of $\La^a$ and $\La^s$) are to the right. Since, in
this form, the
remaining elements only interchange entries within each sectors,
their action will
amount to nothing after the `bar' operation has been performed.  We
can therefore
assume that all the $T_{ij}$'s in
      $T_{i_1 j_1} \cdots T_{i_\ell j_\ell}$
interchange elements between the fermionic and bosonic sectors.

Before going into the proof of the lemma, let us first give an
example that will hopefully shed some
light on the many steps involved in the proof.  Let us consider the
superpartition
\begin{equation}
\La=(7,5,4,3,0;9,6,4,4,2,2,1,1,1) \, ,
\end{equation}
       and act on it with
$T_{1,11}$,
and then with
$T_{4,13}$.
%(also permitted since $\La_4=3>\La_{13}=1$)
We have thus
\begin{equation}
T_{4,13}T_{1,11}\La= T_{4,13}T_{1,11}(7,5,4,3,0;9,6,4,4,2,2,1,1,1)=
(2,5,4,1,0;9,6,4,4,2,7,1,3,1) \, .
\end{equation} The superpartition $\Omega$ is obtained by applying the `bar'
operation:
\begin{equation}
\Omega= \overline{(2,5,4,1,0;9,6,4,4,2,7,1,3,1)}=(5,4,2,1,0;9,7,6,4,4,3,2,1,1)
\, .
\end{equation}
Now, can we conclude directly that $\Omega<\La$? No, because even
though $\La_c$
Bruhat dominates the intermediate composition
$(2,5,4,1,0,9,6,4,4,2,7,1,3,1)$, the latter does not dominate (it is
actually dominated by)
the composition $\Omega_c$ associated to the resulting superpartition $\Omega$.
In the intermediate step, we somehow ended up too low to apply a chain of
Bruhat dominance.  This simply indicates that  the `bar' operation is not
compatible with the ordering on compositions. Therefore,
the
lemma does not follow immediately from the
previous property.  Actually, what the proof of the lemma gives is a precise
construction to arrive at $\Omega$ via a sequence of $T_{ij}$'s without
introducing rearrangements at any intermediate step.

\bigskip

\n {\it Proof:}   \quad Essentially, we want to show that any
$\Omega$ that can be obtained from $\Lambda$
by exchanging a certain number of elements of
       $\La^a$
and   $\La^s$,  and then rearranging both vectors, can also be obtained
       by simply applying a sequence of $T_{ij}$'s, without rearrangement.
      Let $(a_1,\dots,a_\ell)$ be the
partition corresponding to the elements of $\La^a$ that will be moved to the
symmetric side. Also, let $(p_1,\dots,p_\ell)$ be their respective
positions in $\La$, and
$(p'_1,\dots,p'_\ell)$ be their final positions, that is, their positions in
$\Omega$.
Similarly, let
$(b_1,\dots,b_\ell)$ be the partition corresponding to the elements of $\La^s$
that will be moved to the antisymmetric side, and denote by
$(q_1,\dots,q_\ell)$
       their positions in $\La$, and by
$(q'_1,\dots,q'_\ell)$ their final positions in
$\Omega$.
       Because we move larger elements
to the symmetric side, we must have
$a_k > b_k$ for all $k=1,...,\ell$. In our example, we have $\ell=2$, and
\begin{eqnarray}
&(a_1,a_2)=(7,3)\qquad (p_1,p_2)=(1,4)\qquad (p_1',p_2')= (7,11)\\
&(b_1,b_2)=(2,1)\qquad (q_1,q_2)=(10,12)\qquad (q_1',q_2')= (3,4)\;.
\end{eqnarray}
       Now, start from $\Lambda$ and move $(a_1,\dots,a_\ell)$ so that they
occupy the intermediate  positions $q'_1,\dots,q'_{\ell}$
respectively.  This can be
done using a sequence of $T_{ij}$ because, from $a_k > b_k $, we know
that all the
$a_k$'s are moved to the right passed smaller elements. The precise sequence of
$T_{ij}$'s that performs this operation is $T_{p_1 q'_1} \cdots
T_{p_{\ell} q'_{\ell}}$.  In the resulting vector, move
$(b_1,\dots,b_\ell)$ so  that they occupy positions $q_1,\dots,q_{\ell}$
respectively.  Again this can be done using $T_{ij}$ operators because, from
$a_k > b_k$ and  choosing $b_{m}$ such that it occupies the leftmost position
whenever  there
are multiplicities, all the $a_k$'s will be moved to the left passed
larger elements. This amounts to applying $T_{p'_1 q_1} \cdots
T_{p'_{\ell} q_{\ell}}$. Finally, applying the sequence $T_{p'_1
q'_1}
\cdots  T_{p'_{\ell} q'_{\ell}}$ gives $\Omega$.  Transposing these various
steps to our example yields
\begin{equation}
T_{3,7}T_{4,11}T_{7,10}T_{11,12}T_{1,3}T_{4,4}(7,5,4,3,0;9,6,4,4,2,2,1,1,1)
=(5,4,2,1,0;9,7,6,4,4,3,2,1,1)= \Omega
\end{equation}
This shows that $\Omega\leq \La$.
       \hfill $\square$

\begin{corollary} \label{important}
Let $\mu=(\mu_1,\dots,\mu_m;\mu_{m+1},\dots,\mu_N)$ be such
that $\overline{\mu}=\Omega$.  If $\Omega \leq \Lambda$, then
$\overline{T_{ij} \, \mu} \leq \Lambda$.
\end{corollary}
\n {\it Proof:} \quad Since $\overline{\mu} =\Omega$,
$\mu$ can be written
as $\mu = T_{i_1 j_1} \cdots T_{i_{\ell} j_{\ell}} \Omega$, for some
operators $T_{i_{k} j_k}$, $k=1,\dots,\ell$.  Therefore,
from
Lemma~\ref{lempasimpo},
\begin{equation}
\overline{T_{ij} \, \mu}=  \overline{ T_{ij} \,
T_{i_1 j_1} \cdots T_{i_{\ell} j_{\ell}} \Omega}
\leq \Omega \, ,
\end{equation}
which gives $\overline{T_{ij} \, \mu} \leq \Lambda$ if
$\Omega \leq \Lambda$.
\hfill $\square$

\section{Triangular Operators and Determinants } \label{secintro}

This section presents basic results regarding triangular operators.
We should point out that Theorem~\ref{deter} and Corollary~\ref{lrr:cor}
appear for instance in a disguised form in \cite{Mac}.
The exposition of the material in this section follows that of
\cite{FLJ,LLM}.

Let $\{s_{\Lambda}\}_{\Lambda}$ be any basis of $P^{S_N}$.
We write $P^{(s)}_{\Lambda,\preceq}$ for the finite-dimensional
subspace of $P^{S_{N}}$ spanned by the $s_{\Omega}$'s
such that $\Omega \preceq \Lambda$, with
respect to some
ordering
$\preceq$ (which could be any of the three orderings introduced previously),
{\it i.e.},
\begin{equation}
P^{(s)}_{\Lambda,\preceq}=\text{Span}\{
s_\Omega\}_{ \Omega\preceq\Lambda}\,.
\end{equation}
\begin{definition}
A linear operator $O_t:P^{S_N}\to  P^{S_N}$
is called {\em
triangular} if $O_t(P^{(s)}_{\Lambda,\preceq})\subseteq
P^{(s)}_{\Lambda,\preceq}$ for every superpartition $\Lambda$.
\end{definition}

The triangularity of a linear operator $O_t$ in $P^{S_N}$
reduces its eigenvalue problem to a finite-dimensional one.
Triangular operators
can be diagonalized through a determinantal representation of the
eigenfunctions. The triangularity implies that the
expansion of $O_t \, s_\Lambda$
is of the form
\begin{equation}\label{Daction}
O_t\, s_\Lambda = \epsilon_{\Lambda} \, s_{\Lambda}+\sum_{\Omega\prec\Lambda}
d_{\Lambda \Omega}\, s_\Omega  ,
\end{equation}
with the diagonal matrix elements $\epsilon_\Lambda$
being precisely the eigenvalues of $O_t$.

\begin{definition}
The triangular operator $O_t$ is called {\em regular} if
$\epsilon_\Omega\neq\epsilon_\Lambda$ whenever $\Omega \prec \Lambda$.
\end{definition}
Let
$\{ p_\Lambda \}_{\Lambda}$ be a corresponding
basis of eigenfunctions diagonalizing $O_t$. Clearly, we can choose
$p_\Lambda$ to have an expansion of the form
\begin{equation}\label{mexp}
p_\Lambda = s_{\Lambda} + \sum_{\Omega \prec\Lambda}
c_{\Lambda \Omega}\, s_\Omega \, ,
\end{equation}
where the normalization has been chosen to make $p_\Lambda$ monic. The
following theorem provides an explicit determinantal formula for
$p_\Lambda$, given the action of $O_{t}$ on $s_\Lambda$  expressed in
the basis $s_\Lambda$.

\begin{theorem}\label{deter}
Let $O_t$ be a regular triangular operator in $P^{S_N}$ whose
action on the basis $\{s_{\Lambda} \}_{\Lambda}$ is given by
\eqref{Daction}.
Then the unique monic basis $\{
p_\Lambda\}_{\Lambda}$ of $P^{S_N}$ triangularly related to the basis
$\{s_{\Lambda} \}_{\Lambda}$
      (cf. \eqref{mexp}) diagonalizing $O_t$, {\it i.e.,}
\begin{equation}
O_t \,  p_\Lambda =\epsilon_\Lambda\, p_\Lambda ,\;\;\;\;\; \forall
\, \Lambda,
\end{equation}
is given explicitly by the (lower) Hessenberg determinant
         \begin{equation}
         p_\Lambda =\frac{1}{\mathcal{E}_\Lambda}
         \begin{vmatrix}
         s_{\Lambda^{(1)}} & \epsilon_{\Lambda^{(1)}}-\epsilon_{\Lambda^{(n)}}
         & 0 & \hdots & \hdots &
0 \\
         s_{\Lambda^{(2)}} &  d_{\Lambda^{(2)}\Lambda^{(1)}}&
\epsilon_{\Lambda^{(2)}}
-\epsilon_{\Lambda^{(n)}} &
          0 &\hdots & 0 \\
\vdots & \vdots & & \ddots & \ddots & \vdots \\
\vdots & \vdots & \vdots & & \ddots & 0 \\
\makebox[1ex]{} s_{\Lambda^{(n-1)}} &
d_{\Lambda^{(n-1)}\Lambda^{(1)}}& d_{\Lambda^{(n-1)}\Lambda^{(2)}}
&\cdots & &
\epsilon_{\Lambda^{(n-1)}}-\epsilon_{\Lambda^{(n)}} \\
s_{\Lambda^{(n)}} &  d_{\Lambda^{(n)}\Lambda^{(1)}}&
d_{\Lambda^{(n)}\Lambda^{(2)}} & \cdots
        &\cdots &
d_{\Lambda^{(n)}\Lambda^{(n-1)}}
         \end{vmatrix} .
         \end{equation}
Here $\Lambda^{(1)}<\Lambda^{(2)}<\cdots
<\Lambda^{(n-1)}<\Lambda^{(n)}=\Lambda$ denotes any linear
ordering, refining the natural order $\preceq$, of the superpartitions,
$\Lambda^{(i)}, i=1,\dots,n-1$,
that precede $\Lambda$ in the ordering $\preceq$.
The
normalization is determined by
\begin{equation}
\mathcal{E}_\Lambda =
\prod_{i=1}^{n-1}
(\epsilon_{\Lambda}-\epsilon_{\Lambda^{(i)}}) \, .
\end{equation}
\end{theorem}

With $O_t=\H_2$ and $s_\La=m_\La$, the previous theorem leads to a closed
expression
for the $\H_2$ eigenfunctions, the Jack superpolynomials
$\J_\La$ of \cite{DLM1,DLM2}, in terms of the coefficients
$d_{\La\Omega}$ entering
in the supermonomial decomposition of $\H_2 \, m_\La$. These
coefficients have been
computed in
\cite{DLM2}. As already indicated, the superpolynomials $\J_\La$ are
not orthogonal.
We will seek  linear combinations that are orthogonal by considering
the eigenfunctions of $\I_1$. Theorem \ref{deter} will then be invoked
again, but this
time with $s_\La=\J_\La$ and $O_t=\I_1$. Computing the action of $\I_1$ in the
$\J_\La$ basis will provide closed form formulas for the orthogonal
superpolynomials $J_\La$.  In the $m_\La$ basis,  $J_\La$ will appear
as a
determinant of determinants.

As an aside, we point out that the determinantal formula for $p_\Lambda$,
leads to a linear recurrence relation encoding an efficient
algorithm for the computation of the coefficients $c_{\Lambda \Omega}$
entering the expansion \eqref{mexp}.

\begin{corollary} \label{lrr:cor}
The expansion of $p_\Lambda$ is of the form
\begin{equation}
p_\Lambda = \sum_{\ell=1}^{n} c_{\Lambda\Lambda^{(\ell)}}\,
s_{\Lambda^{(\ell)}},
\end{equation}
with $c_{\Lambda\Lambda^{(n)}}=c_{\Lambda \Lambda}=1$ and, for
$1<\ell \leq n$,
\begin{equation}
        c_{\Lambda\Lambda^{(\ell-1)}}
=\frac{1}{\epsilon_\Lambda-\epsilon_{\Lambda^{(\ell-1)}}}
\sum_{k=\ell}^n c_{\Lambda\Lambda^{(k)}} \,
d_{\Lambda^{(k)}\Lambda^{(\ell-1)}} \, .
\end{equation}
\end{corollary}

We conclude this section with an elementary and surely well known
proposition that we prove for a lack of reference.
It provides a
simple way of
computing the $p_\la$ eigenvalues of mutually commuting operators in terms of
the action of these operators on the $s_\La$ basis.

\begin{proposition} \label{propDt}
       Let $D_t$ be a triangular operator commuting with
$O_t$.  Then,
\begin{equation}
D_t p_{\Lambda} = \varepsilon_{\Lambda} p_{\Lambda} \, ,
\end{equation}
where $\varepsilon_{\Lambda}$ is the coefficient of $s_{\Lambda}$ in
$D_t s_{\Lambda}$.
\end{proposition}
\noindent {\it Proof.} \quad Let $\tilde p_{\Lambda}=D_t
p_{\Lambda}/\varepsilon_{\Lambda}$.  Then, from \eqref{mexp} and the fact that
$D_t$ is a
triangular operator, $\tilde p_{\Lambda}$ is seen to be of the form
\begin{equation} \label{sexp}
\tilde p_{\Lambda} = s_{\Lambda} + \sum_{\Omega \preceq \Lambda}
g_{\Lambda \Omega} \, s_{\Omega} \, .
\end{equation}
Now, because $O_t$ and $D_t$ commute, we have
$O_t \tilde p_{\Lambda}= D_t O_t p_{\Lambda}/\varepsilon_{\Lambda}=
\epsilon_{\Lambda} \tilde p_{\Lambda}$.  Therefore, the monic polynomial
$\tilde p_{\Lambda}$
diagonalizes $O_t$ and, from \eqref{sexp}, is triangularly related to the
basis $\{s_{\Lambda} \}_{\Lambda}$.  From the uniqueness in
Theorem~\ref{deter}, we must have $\tilde p_{\Lambda}=p_{\Lambda}$, or
$D_t p_{\Lambda} =
\varepsilon_{\Lambda} p_{\Lambda}$.
\hfill $\square$

\section{The action of  ${\mathcal H}_n$}

We are now ready to tackle one of our main objectives,
which is to obtain common
eigenfunctions of the commuting operators ${\mathcal H}_n$ and
${\mathcal I}_n$. In
this section, we first study the action of the
${\mathcal H}_n$'s.

We start with a very simple proposition concerning the operators
${\mathcal O}_{ij}$ that we state without proof.

\begin{proposition} \label{propO}
        If we only consider terms that are permutations of
$z_i^{\lambda_i} z_j^{\lambda_j}$, we have, for $i>j$,
\begin{equation}
{\mathcal O}_{ij} \, z_i^{\lambda_i} z_j^{\lambda_j} =
\begin{cases}
\phantom{-}z_i^{\lambda_i}z_j^{\lambda_j}  & \lambda_i > \lambda_j \\
-z_i^{\lambda_j}z_j^{\lambda_i}  & \lambda_i < \lambda_j \\
\phantom{-}0 & {\rm otherwise}
       \end{cases} \, ,
\end{equation}
and, for $i<j$,
\begin{equation}
{\mathcal O}_{ij} \, z_i^{\lambda_i} z_j^{\lambda_j} =
\begin{cases}
\phantom{-}z_i^{\lambda_j}z_j^{\lambda_i}  & \lambda_i > \lambda_j \\
-z_i^{\lambda_i}z_j^{\lambda_j}  & \lambda_i < \lambda_j \\
\phantom{-} 0 & {\rm otherwise}
       \end{cases} \, .
\end{equation}
\end{proposition}

\begin{lemma} \label{lemtril} Let $\lambda$ be a partition, and let
$\lambda^R$ be $\lambda$ in reverse order.
        Then
\begin{equation} \label{triH}
{\mathcal H}_n \, z^{\lambda^R} = \varepsilon_{n,\lambda}
\, z^{\lambda^R} +
\sum_{\mu < \lambda^{R}; \mu^+ \neq \lambda} a_{n,\mu} \, z^{\mu} \, ,
\end{equation}
with $\varepsilon_{n,\lambda}$ given explicitly by the formula
\begin{equation}
\varepsilon_{n,\lambda}  = \sum_{i=1}^N \left(
\underline{ \lambda_i^R }\right)^n \, ,
\end{equation}
where the symbol $\underline{\gamma_i}$
was introduced in Property~\ref{proptriDi}, and  where $\la^R_i$ stands
for $(\la^R)_i$.
\end{lemma}
\noindent{\it Proof.} \quad The lemma will hold if we can demonstrate that, for
       $\lambda_1 \geq \lambda_2 \geq \cdots \geq \lambda_N$,
we have
\begin{equation}
\label{Di}
{\mathcal D}_i \, z^{\lambda^{R}} = \underline{{ \lambda}_i^R} \,
z^{\lambda^{R}} +
\sum_{\mu< \lambda^{R}; \mu^+ \neq \lambda}
       a_{\mu} \, z^{\mu} \, .
\end{equation}
Using Property~\ref{proptriDi}, for this to be true we only need to show that
terms of the type $z^{\mu}$,
where $\mu^+=\lambda$ never occur (except for $z^{\lambda^R}$).
Since $\lambda^R \leq \mu$ for any $\mu$ such that $\mu^+=\lambda$, this
is indeed seen to be true.
\hfill $\square$

The special action of ${\mathcal H}_n$ on $z^{\lambda^R}$ induces
a  triangularity on $m_{\Lambda}$.
\begin{theorem}  Let $\Lambda^*=(\lambda_1,\dots,\lambda_N)=\lambda$.  Then,
\begin{eqnarray}
\label{trihno}
{\mathcal H}_n \, m_{\Lambda}& =& \varepsilon_{n,\lambda}
\, m_{\Lambda} +
\sum_{\Omega <_h \Lambda} a_{\Lambda\Omega}^{(n)} \, m_{\Omega}  \, ,
\end{eqnarray}
with $ \varepsilon_{n,\lambda}$ given in Lemma~\ref{lemtril}.
\end{theorem}
\noindent{\it Proof.}  \quad We have that $\theta_{\{1,\dots,m \}}
z^{\Lambda}=
\pm {\mathcal K}_{\sigma'} \, \theta_{I} z^{\lambda^R}$, for some
$\sigma' \in S_N$
and some $I \subseteq \{1,\dots,N \}$.  Since ${\mathcal H}_n$ commutes with
${\mathcal K}_{ij}$ and $\theta_I$, we therefore obtain, using equation
\eqref{mono} and Lemma~\ref{lemtril},
\begin{eqnarray}
{\mathcal H}_n \, m_{\Lambda}
& = & \pm   \frac{1}{f_{\Lambda}} \sum_{\sigma \in S_N}
{\mathcal K}_{\sigma} {\mathcal K}_{\sigma'}
\, \theta_I {\mathcal H}_n \, z^{\lambda^R} \nonumber \\
& =&  \pm  \frac{1}{f_{\Lambda}} \sum_{\sigma \in S_N} {\mathcal K}_{\sigma}
{\mathcal K}_{\sigma'} \theta_I
        \left(
\varepsilon_{n,\lambda}
\, z^{\lambda^R} +
\sum_{\mu < \lambda^R;\mu^+ \neq \lambda}  a_{n, \mu}
\, z^{\mu} \right)  \nonumber \\
&=&  \varepsilon_{n,\lambda}
\, m_{\Lambda} +
\sum_{\Omega <_h \Lambda} a_{\Lambda\Omega}^{(n)} \, m_{\Omega} \, ,
\nonumber
\end{eqnarray}
which proves the theorem.
\hfill $\square$

The next theorem was proven in \cite{DLM2} (cf. Theorem 10 therein).
\begin{theorem}
\label{theo1}   There exists a unique basis
$\{ {\mathcal J}_{\Lambda} \}_{\Lambda}$ of $P^{S_N}$
such that
\begin{equation}
        {\mathcal H}_2 {\mathcal J}_{\Lambda} = \varepsilon_{\lambda,2}
\, {\mathcal J}_{\Lambda}  \qquad {\rm and}\qquad {\mathcal J}_{\Lambda} =
m_{\Lambda} + \sum_{\Omega <_h \Lambda} v_{\Lambda \Omega} \, m_{\Omega} \, ,
\end{equation}
where $\lambda=\Lambda^*$.
\end{theorem}
The explicit action of $\H_2$ on the supermonomial basis, $m_{\La}$,
was computed in \cite{DLM2}.  In view of Theorem~\ref{deter}, a determinantal
formula giving ${\mathcal J}_{\Lambda}$ in terms of supermonomials
immediately followed.  It should be noted that an efficient algorithm
to evaluate such a determinant can be found in Corollary~\ref{lrr:cor}.

We now show that ${\mathcal J}_{\Lambda}$ is an eigenfunction of $\H_n$,
for all $n$.
\begin{theorem} \label{theo2} With $\Lambda^*=\lambda$,
we have
\begin{equation}
{\mathcal H}_n \, {\mathcal J}_{\Lambda} =  \varepsilon_{\lambda,n} \,
{\mathcal J}_{\Lambda}
=\left[ \sum_{i=1}^N (\underline{ \lambda_i^R})^n \right] {\mathcal
J}_{\Lambda} \, .
\end{equation}
Furthermore, for $\mu$ a partition such that
$\mu \neq \lambda$, there exists at least
one $n$ such that
        $\varepsilon_{\lambda,n} \neq
\varepsilon_{\mu,n}$.
\end{theorem}
\noindent {\it Proof.} Given Theorem~\ref{theo1}, formula
\eqref{trihno} and the fact that
${\mathcal H}_n$ and ${\mathcal H}_2$ commute,
the first part of the theorem follows immediately
from Proposition~\ref{propDt}.
The second part of the theorem
is obvious because the
        $\varepsilon_{\lambda,n}$ ($n=1,2,\dots$) are
        polynomials
in $\beta$ whose constant terms $\varepsilon_{\lambda,n} \big|_{\beta=0}
= \sum_{i=1}^N \lambda_i^n$
are such that if
$\varepsilon_{\lambda,n} \big|_{\beta=0}=
\varepsilon_{\mu,n} \big|_{\beta=0}$  for all $n=1,2,\dots$,
then $\lambda=\mu$. Therefore, $\varepsilon_{\lambda,n}$ and
$ \varepsilon_{\mu,n}$,
considered as functions of a generic parameter $\beta$, cannot be equal.
\hfill $\square$

The last theorem implies that the superpolynomials $\J_\La$ and
$\J_\Omega$ associated to distinct superpartitions such that
$\La^*=\Omega^*$, share the same eigenvalues. Therefore, additional commuting
operators need to be diagonalized in order to lift the degeneracies. These are
the $\I_n$ charges, whose action is considered in  the following section.

\section{The action of ${\mathcal I}_n$}

Let $\I'_n$ be the operator
\begin{equation}
\I'_n = \sum_{i=1}^N \K_{1 i} \, \theta_1 \frac{\d}{\d \theta_1} \, \D_1^n \, .
\end{equation}
The following proposition states that $\I_n$ and $\I'_n$ are
equivalent (up to a constant) on $P^{S_N}$.
\begin{proposition} \label{IIp} We have, for $f \in P^{S_N}$,
\begin{equation}
\I_n \, f = (N-1)! \, \I'_n \, f \, .
\end{equation}
\end{proposition}

\n {\it Proof:} \quad The symmetric group can be factorized in the
following way
\begin{equation}
\sum_{\sigma \in S_N} \K_{\sigma} = \sum_{i=1}^N \K_{{1i}}
\sum_{w \in S_{\{2,\dots,N \}}} \K_{w} \, .
\end{equation}
Therefore, since  for $w \in S_{\{2,\dots,N \}}$, $\K_w$
commutes with $\theta_1 \frac{\d}{\d \theta_1} \, \D_1^n$, and since
       $\K_{w}$  leaves $f$ invariant, the proposition follows.
\hfill $\square$

We now introduce a subspace of $P^{S_N}$.
\begin{definition}
For $\Lambda$ a superpartition in the $m$-fermion sector,
       $L_{\Lambda}$ is given by
$$
L_{\Lambda} = {\rm Span} \left \{ z^{\mu} \, \big|
\, \overline{\mu}
       \leq \Lambda
         \right \} = {\rm Span} \left \{ K_{\sigma} K_{w} z^{\Omega} \, \big|
\, \sigma \in S_m, w \in S_{\{m+1,\dots,N \}} , {\rm and~} \Omega
       \leq \Lambda
         \right \} \, ,
$$
where as usual $\Omega$ is a superpartition.
\end{definition}
This subspace has the following property.
\begin{lemma} \label{Llambda}
       Let $\Lambda$ be a superpartition in the $m$-fermion
sector, and let $i \in \{1,\dots,m\}$. Then,
\begin{equation}
\D_i \left( L_{\Lambda} \right) \subseteq
L_{\Lambda} \, .
\end{equation}
\end{lemma}
\noindent{\it Proof.} From the action of $\D_i$ given in
Property~\ref{proptriDi},
the lemma is false only if,  for $\bar \mu \leq \Lambda$,
there exist, in $\D_i \, z^{\mu}$,
some terms of the type $z^{\nu}$, where $\nu^+=\Lambda^*$ and
$\nu \not \leq \Lambda$.  Given that by definition, $\mathcal D_i$
decomposes into the blocks ${\mathcal O}_{ij}$, it
is thus sufficient to limit ourselves to the terms of
${\mathcal O}_{ij} z^{\mu}$ considered in
Proposition~\ref{propO}.  If
$j \in \{ 1,\dots,m\}$,
these terms are seen to be
of the type $z^{\nu}$, with
$\bar \nu= \bar \mu \leq \Lambda$, and hence belong to $ L_{\Lambda}$.
On the other
hand, when $j \in \{ m+1,\dots,N\}$, since $i \in  \{ 1,\dots,m\} $, we are
always in the case $j>i$ of Proposition~\ref{propO},
in which case these terms are of the type $z^{T_{ij} \mu}$.
      The lemma then follows because,
from Corollary~\ref{important}, $\overline{T_{ij} \mu} \leq \Lambda$
whenever $\bar{\mu} \leq \Lambda$.
\hfill $\square$

\begin{theorem}  We have
\begin{equation} \label{trii}
{\mathcal I}_n \, m_{\Lambda} =  \epsilon_{\Lambda,n}
        \, m_{\Lambda} +
\sum_{\Omega < \Lambda} b_{\Lambda \Omega}^{(n)} \, m_{\Omega} \, .
\end{equation}
\end{theorem}
\n {\it Proof:} \quad
We will prove the  equivalent statement that $\I'_n$ (see
Proposition~\ref{IIp}) acts triangularly. We have
\begin{equation}
\I'_n \, m_{\Lambda} = \sum_{i=1}^N \K_{1 i} \,
\theta_1 \frac{\d}{\d \theta_1} \, \D_1^n \frac{1}{f_{\Lambda}}
\sum_{w \in S_N} \K_w \, \theta_{\{1,\dots,m \}} z^{\Lambda} \, .
\end{equation}
We will now focus on the part of this expression involving
$\theta_{\{1,\dots,m \}}$ to the left.  This term is of the form
\begin{equation} \label{thetaterm}
\I'_n \, m_{\Lambda} \Big|_{\theta_{\{1,\dots,m \}}} =
\frac{1}{f_{\Lambda}} \sum_{w \in S_m}
(-1)^{\ell(w)+1} \sum_{i=1}^m K_{1 i}
       \, \D_1^n \, z^{w\Lambda} \, ,
\end{equation}
where $\ell(w)$ is the length of the permutation $w$. This is
because, if $\kappa_w
\theta_{\{1,\dots,m
\}}$ does not contain
$\theta_1$, the term will be annihilated by $\frac{\d}{\d \theta_1}$.
Now, if $\kappa_w \theta_{\{1,\dots,m \}}$ contains $\theta_1$,
then $\kappa_{1 i} \kappa_w \theta_{\{1,\dots,m \}}$ contains $\theta_i$,
and thus $\kappa_{1 i} \kappa_w \theta_{\{1,\dots,m \}}$
will certainly not be equal to $\pm \theta_{\{1,\dots,m \}}$ if $i >m$.
Therefore $i \leq m$, which means that $w$ needs to belong to $S_m$ for
$\kappa_{1 i} \kappa_w \theta_{\{1,\dots,m \}}$ to be equal to
$\pm \theta_{\{1,\dots,m \}}$. Finally, with
$\kappa_{1 i} \kappa_w \theta_{\{1,\dots,m \}}= (-1)^{\ell(w)+1}
\theta_{\{1,\dots,m \}}$, formula \eqref{thetaterm} is seen to hold.

Now, if $w \in S_m$, then $z^{w \Lambda} \in L_{\Lambda}$, which implies, from
Lemma~\ref{Llambda}, that $\D_i^n z^{w \Lambda} \in L_{\Lambda}$. Therefore,
since
$K_{1i}$, for $i=1,\dots,m$, also preserves $L_{\Lambda}$ (by definition),
we have that $\I'_n \, m_{\Lambda} |_{\theta_{\{1,\dots,m \}}}$
belongs to $L_{\Lambda}$.  Therefore, any $z^{\Omega}$ in
$\I'_n \, m_{\Lambda} |_{\theta_{\{1,\dots,m \}}}$, for $\Omega$ a
superpartition, will be such that $\Omega \leq \Lambda$, which
proves the theorem.
\hfill $\square$

Note that the eigenvalues $\epsilon_{\Lambda,n} $ are not given explicitly.
However, $\epsilon_{\Lambda,1}$ is obtained in the next theorem.  This
theorem also characterizes the precise action of $\I_1$ on
monomials, if we discard  coefficients that will not be needed in the
sequel.
\begin{theorem}  \label{bigtheo} The action of ${\mathcal I}_1$ on the
monomials is  the following
\begin{equation} \label{Ionm}
{\mathcal I}_1 \, m_{\Lambda} = \epsilon_{\Lambda,1} \, m_{\Lambda}
+ \sum_{\Omega <_t \Lambda } b_{\Lambda \Omega}
        \, m_{\Omega}
+  \sum_{\Gamma <_{h} \Lambda }
c_{\Lambda \Gamma} \, m_{\Gamma} \, ,
\end{equation}
with
\begin{equation}
\epsilon_{\Lambda,1} = (N-1)! \left[ \sum_{i=1}^m \Lambda_i - \beta
\Bigl( m(m-1) +
\#_{\Lambda}
        \Bigr)  \right] \, ,
\end{equation}
where $\#_{\Lambda}$ is the number of
pairs $(i,j)$ such that
$i \in \{1,\dots,m\}$, $j \in \{m+1,\dots,N \}$ and $\Lambda_i <
\Lambda_j$, and
with
\begin{equation}
b_{\Lambda \Omega}=
\begin{cases}
\label{eqb} (N-1)! \, \beta \,  {\rm sgn}(\sigma^a_{T_{ij}
\Lambda} ) \, n_{\Omega^s}(\Lambda_i) &
{\rm{if~}} \Omega = \overline{T_{ij} \Lambda} {\rm ~~for~some~~} i<j  \\
0 & {\rm otherwise}
\end{cases} \,
\end{equation}
(the $ c_{\Lambda \Gamma}$'s are left undetermined).
\end{theorem}

The action of $\I_1$ given in (\ref{Ionm}) displays two types of subleading
terms, each type being characterized by one of the two specializations of the
Bruhat ordering. As in the action of $\H_n$, we recover terms that are
$h$-ordered. But in addition, there appear terms that are $t$-ordered,
labeled by superpartitions such that
$\La^*=\Omega^*$. It is precisely the superpolynomials associated to such
superpartitions that were $\H_n$-degenerate. It is because they
can now be compared that the action of $\I_1$ lifts the degeneracies, as
we will see in the lemma that follows this theorem.

\noindent {\it Proof.} \quad To simplify the analysis, we again work with
$\I'_1$ instead of $\I_1$, and focus on the coefficient
$\theta_{\{1,\dots,m \}}$.  From \eqref{thetaterm}, this coefficient is given
by
\begin{equation} \label{bigthetaterm}
\I'_1 \, m_{\Lambda} \Big|_{\theta_{\{1,\dots,m \}}} =
\frac{1}{f_{\Lambda}} \sum_{i=1}^m K_{1 i}
       \, \D_1 \sum_{w \in S_m}
(-1)^{\ell(w)+1}  \, z^{w\Lambda} \, .
\end{equation}
      From the Hecke algebra relations,
and $(1i) = \sigma_{i-1} \cdots \sigma_1 \cdots \sigma_{i-1}$
we obtain
\begin{equation}
       K_{1 i}  \, \D_1 = \D_i \,  K_{1 i} + \beta \sum_{j=1}^{i-1}
K_{(\sigma_{i-1} \cdots \sigma_1)_j \,  \sigma_2 \cdots \sigma_{i-1}} \, ,
\end{equation}
where the symbol $()_j$ means that $\sigma_j$ does not belong to the product
in parenthesis.  Now, the transposition $(1i)$ contains an odd number of
elementary transpositions, while all the terms of the form
$(\sigma_{i-1} \cdots \sigma_1)_j \,  \sigma_2 \cdots \sigma_{i-1}$ contain
an even number of such transpositions.  With $\sum_{w \in S_m}
(-1)^{\ell(w)+1}  \, z^{w\Lambda}$ being totally antisymmetric in the first
$m$ variables, we thus
obtain
\begin{eqnarray} \label{newbigthetaterm}
\I'_1 \, m_{\Lambda} \Big|_{\theta_{\{1,\dots,m \}}} & = &
\frac{1}{f_{\Lambda}} \sum_{i=1}^m \Bigl(
       \D_i -\beta(i-1) \Bigr) \sum_{w \in S_m}
(-1)^{\ell(w)}  \, z^{w\Lambda} \nonumber \\
& = &
\frac{1}{f_{\Lambda}}
\sum_{w \in S_m}
(-1)^{\ell(w)} \, K_{w} \sum_{i=1}^m \Bigl(
       \D_i -\beta(i-1) \Bigr) z^{\Lambda} \, ,
\end{eqnarray}
where we have used the fact that $\sum_{i=1}^m \D_i$ commutes with
any $K_{w}$ such that $w \in S_m$.
We have, from \eqref{defDi},
\begin{equation}
       \sum_{i=1}^m \Bigl(
       \D_i -\beta(i-1) \Bigr) = \sum_{i=1}^m z_i \frac{\d}{\d z_i}
+ \beta \sum_{1 \leq i < j \leq m} ({\mathcal O}_{ij}+{\mathcal O}_{ji})
+\beta \sum_{i=1}^m \sum_{j=m+1}^n {\mathcal O}_{ij} -\beta m(m-1) \, .
\end{equation}
With ${\mathcal O}_{ij}+{\mathcal O}_{ji}=0$, this leads to
\begin{equation} \label{almost}
\I'_1 \, m_{\Lambda} \Big|_{\theta_{\{1,\dots,m \}}} =
\left[ \sum_{i=1}^m \Lambda_i -\beta m(m-1) \right] m_{\Lambda} +
\frac{\beta}{f_{\Lambda}} \sum_{w \in S_m}
(-1)^{\ell(w)} \, K_{w} \sum_{i=1}^m \sum_{j=m+1}^N  {\mathcal O}_{ij}
        \, z^{w\Lambda} \, .
\end{equation}
The last term of this expression becomes (if we do not consider coefficients
that are not permutations of $\Lambda$)
\begin{eqnarray}
&~&  \frac{\beta}{f_{\Lambda}}  \sum_{w \in S_m}
(-1)^{\ell(w)} K_{w}
       \sum_{i=1}^m
\sum_{j=m+1}^N  {\mathcal O}_{ij}
       \, z^{\Lambda} \\
&~& \qquad \qquad \qquad = -\frac{\beta \#_{\Lambda}}{f_{\Lambda}}
\sum_{w \in S_m}
(-1)^{\ell(w)} K_{w}
       \, z^{\Lambda} + \frac{\beta}{f_{\Lambda}}  \sum_{w \in S_m}
(-1)^{\ell(w)} K_{w} \sum_{(i,j);\Lambda_i>\Lambda_j}
       \, z^{T_{ij}\Lambda} \, , \nonumber
\end{eqnarray}
where $(i,j)$ is considered to be such that $i \in \{1,\dots,m \}$
and $j \in \{m+1,\dots,N \}$.
Putting everything together, we get
\begin{eqnarray}
{\mathcal I'}_1 \, m_{\Lambda}
& = & \epsilon_{\Lambda,1}\, m_{\Lambda}
        + \frac{\beta}{f_{\Lambda}}  \sum_{w \in S_N} {\mathcal K}_{w} \,
\theta_1 \cdots \theta_m \, \sum_{(i,j) \, ; \, \Lambda_i > \Lambda_j}
z^{T_{ij} \Lambda} + \sum_{\Gamma <_{h} \Lambda }
b_{\Lambda \Gamma} \, m_{\Gamma} \, .
\end{eqnarray}
If $\Omega= \overline{T_{ij} \Lambda}$, the coefficient
of $m_{\Omega}$ in the last formula is then given by
\begin{eqnarray}
\beta \, {\rm sgn}(\sigma^a_{T_{ij} \Lambda} )
\frac{f_{\Omega}}{f_{\Lambda}} n_{\La^s}(\Lambda_j)
& = & \beta \, {\rm sgn}(\sigma^a_{T_{ij} \Lambda} )
\frac{n_{\Omega^s}(\Lambda_i)!n_{\Omega^s}(\Lambda_i)!}{n_{\La^s}
(\Lambda_j)!n_{\La^s}(\Lambda_i)!}
\bigl( n_{\Omega^s}(\Lambda_j)+1 \bigr) \nonumber \\
& = & \beta \, {\rm sgn}(\sigma^a_{T_{ij} \Lambda} )
\frac{n_{\Omega^s}(\Lambda_i)}{n_{\Omega^s}(\Lambda_j)+1}
\bigl( n_{\Omega^s}(\Lambda_j)+1 \bigr) \nonumber \\
& = &   \beta \,
{\rm sgn}(\sigma^a_{T_{ij} \Lambda} ) \,
n_{\Omega^s}(\Lambda_i) \, ,
\end{eqnarray}
since $n_{\La^s}(\Lambda_i) =n_{\Omega^s}(\Lambda_i)-1$ and
$n_{\La^s}(\Lambda_j) =n_{\Omega^s}(\Lambda_j)+1$.
        \hfill $\square$

\begin{lemma} \label{lem4}
        The triangular operator ${\mathcal I}_1$ is regular,
that is, $\epsilon_{\Lambda,1} \neq \epsilon_{\Omega,1}$
if $\Omega < \Lambda$.
\end{lemma}

\noindent {\it Proof.} \quad
      $\Omega<\La$ means that $\Omega$
differs from $\Lambda$ by the application of a sequence of
$T_{ij}$'s, with $i\leq m$
and $j>m$. It follows that $\sum_{i=1}^m\Omega_i<\sum_{i=1}^m\La_i$,
so that the
constant term in $ \epsilon_{\Omega,1}$, viewed as a polynomial in $\beta$, is
strictly smaller than
$\epsilon_{\La,1}|_{\beta=0}$. This readily implies that $\epsilon_{\Omega,1}
\not=
\epsilon_{\Lambda,1}$.
      \hfill $\square$

\begin{lemma} \label{lem5}
        The action of ${\mathcal I}_1$ on ${\mathcal J}_{
\Lambda}$ is triangular with respect to the $t$-ordering, that is,
\begin{equation}
{\mathcal I}_1 \, {\mathcal J}_{\Lambda} = \epsilon_{\Lambda,1}
\, {\mathcal J}_{\Lambda}+
        \sum_{\Omega <_{t} \Lambda}
b_{\Lambda \Omega}  \, {\mathcal J}_{
\Omega} \, ,
\end{equation}
where the coefficients $b_{\Lambda \Omega}$
are given in  \eqref{eqb}.
\end{lemma}
\noindent {\it Proof.} \quad  From
Theorem~\ref{theo1} and because the
order $\leq_h$ is weaker than the Bruhat order $\leq$, we have
\begin{equation} \label{Jm}
{\mathcal J}_{\Lambda} =
        {m}_{\Lambda}+
        \sum_{\Omega < \Lambda \, ; \, \Omega^* \neq \Lambda^*}
v_{\Lambda \Omega}  \, {m}_{
\Omega} \, ,
\end{equation}
and thus, from \eqref{Ionm},
\begin{eqnarray}
{\mathcal I}_1 \, {\mathcal J}_{\Lambda} & = &
        \epsilon_{\Lambda,1}
\, {m}_{\Lambda} + \sum_{\Omega <_t \Lambda} b_{\Lambda \Omega}
        \, m_{\Omega}
+  \sum_{\Gamma < \Lambda \, ; \, \Gamma^* \neq
\Lambda^*} c_{\Lambda \Gamma} \, m_{\Gamma} \nonumber \\
\label{dtrop}
& = & \epsilon_{\Lambda,1}
\, {\mathcal J}_{\Lambda} +
\sum_{\Omega <_t \Lambda } b_{\Lambda \Omega}
        \, {\mathcal J}_{\Omega}
+  \sum_{\Gamma < \Lambda \, ; \, \Gamma^* \neq
\Lambda^*} d_{\Lambda \Gamma} \, {\mathcal J}_{\Gamma} \, ,
\end{eqnarray}
since from \eqref{Jm} we get the inverse relation
\begin{equation}
{m}_{\Lambda} =
        {\mathcal J}_{\Lambda}+
        \sum_{\Omega < \Lambda \, ; \, \Omega^* \neq \Lambda^*}
w_{\Lambda \Omega}  \, {\mathcal J}_{
\Omega} \, .
\end{equation}
It now suffices to show that the coefficients $d_{\Lambda \Gamma}$ in
\eqref{dtrop} do in fact vanish.  Since ${\mathcal I}_1$ commutes with
${\mathcal H}_n$, ${\mathcal I}_1 \, {\mathcal J}_{\Lambda}$ must be
an eigenfunction of ${\mathcal H}_n$ with eigenvalue
$\varepsilon_{\Lambda,n}$ ($n=1,2,\dots$).  From Theorem~\ref{theo2},
the expansion
of ${\mathcal I}_1 \, {\mathcal J}_{\Lambda}$ in terms of
${\mathcal J}_{\Omega}$ can thus only contain terms such that
$\Omega^*=\Lambda^*$.
\hfill $\square$

\section{The orthogonal Jack superpolynomials $J_{\La}$}

\begin{theorem} \label{theo6}
There exists a unique basis $\{ {J}_{\Lambda} \}_{\Lambda}$
of $P^{S_N}$ such that
\begin{equation}
        {\mathcal I}_1 {J}_{\Lambda} = \epsilon_{\Lambda,1}
\, {J}_{\Lambda}  \qquad {\rm and}\qquad {J}_{\Lambda} =
{\mathcal J}_{\Lambda} + \sum_{\Omega <_{t}\La}
u_{\Lambda \Omega} \,
{\mathcal J}_{\Omega} \, .
\end{equation}
\end{theorem}
\noindent {\it Proof.} \quad Using Lemmas~\ref{lem4} and \ref{lem5}, the
        theorem follows immediately from Theorem~\ref{deter}.
\hfill $\square$

      From Theorem \ref{deter},  Lemma \ref{lem5} and the explicit
expression of the eigenvalues $\varepsilon_{\Lambda,1}$ given in Theorem
\ref {bigtheo}, a determinantal formula, giving
$J_{\La}$ in terms
of $\J_{\Omega}$, can be obtained.  Moreover, by Corollary~\ref{lrr:cor},
a recurrence is provided for the
coefficients
$u_{\Lambda\Omega}$.  Note that because the eigenvalues $
\epsilon_{\Lambda,1}$ and the
coefficients $b_{\La,\Omega}$ given in Theorem \ref {bigtheo} do not
depend upon $N$,
apart from a factorizable overall prefactor $(N-1)!$, the coefficients
$u_{\Lambda\Omega}$ are
$N$-independent. Moreover, since the expansion of $\J_\la$ in the
supermonomial basis
is $N$-independent, this holds true for the supermonomial
decomposition of $J_\la$. This
is illustrated in Appendix A.

Given Theorem~\ref{theo2}, the previous theorem has the following corollary.
\begin{corollary} \label{coro}
\begin{equation}
{\mathcal H}_n \, {J}_{\Lambda} =  \varepsilon_{\Lambda,n} \,
{J}_{\Lambda} \, .
\end{equation}
Furthermore,
if $\Lambda^*\neq \Omega^*$, then there exists at least one
$n$ such that
        $\varepsilon_{\Lambda,n} \neq
\varepsilon_{\Omega,n}$.
\end{corollary}
We now show that $J_{\La}$ is also an eigenfunction of $\I_n$, for all
$n$.
\begin{theorem} \label{theo7}
We have
\begin{equation} \label{InP}
{\mathcal I}_n \, {J}_{\Lambda} =  \epsilon_{\Lambda,n} \,
{J}_{\Lambda} \, .
\end{equation}
Furthermore,
if $\La^a \neq \Omega^a$, then there exists
at least one $n$ such that
        $\epsilon_{\Lambda,n} \neq
\epsilon_{\Omega,n}$.
\end{theorem}
\noindent {\it Proof.} \quad Given Theorem~\ref{theo6}, formula \eqref{trii}
and the fact that ${\mathcal I}_n$ and ${\mathcal I}_1$ commute, the first part
of the theorem follows immediately
from Proposition~\ref{propDt}.
The second part of the  theorem
is
obvious because the $\epsilon_{\Lambda,n}$'s ($n=1,2,\dots$) are
polynomials
in $\beta$ whose constant terms $\epsilon_{\Lambda,n}
\big|_{\beta=0}
= \sum_{i=1}^m \Lambda_i^n$ are such that
if
$\epsilon_{\Lambda,n} \big|_{\beta=0}=
\epsilon_{\Omega,n}
\big|_{\beta=0}$ for $n=1,2,\dots$,
then $\La^a = \Omega^a$. Hence,
$\epsilon_{\La,n}$ and
$ \epsilon_{\Omega,n}$,
considered as
functions of a generic parameter $\beta$, cannot be equal when $\La^a \neq
\Omega^a$.
\hfill
$\square$

Now, it is obvious that if $\Lambda^*=\Omega^*$
and
$\La^a=\Omega^a$, then $\Lambda=\Omega$.  Therefore,
using
Corollary~\ref{coro},
the previous theorem has the
following
corollary.
\begin{corollary} The polynomial ${J}_{\Lambda}$
is the
unique common eigenfunction of the operators
${\mathcal H}_n$
and ${\mathcal I}_\ell$ ($n,\ell = 1,2,\dots$),
with respective
eigenvalues $\varepsilon_{\Lambda,n}$
and
$\epsilon_{\Lambda,\ell}$.
\end{corollary}
We thus have
immediately, since the operators ${\mathcal H}_n$ and
${\mathcal I}_\ell$, $n,\ell = 1,2,\dots$, are self-adjoint
with
respect to the scalar product
$\langle.,. \rangle_{\beta}$, the
orthogonality of the basis
$\{ {J}_{\Lambda}
\}_{\Lambda}$.
\begin{theorem} \label{theoortho}
The basis $\{
{J}_{\Lambda} \}_{\Lambda}$
of $P^{S_N}$
satisfies
\begin{equation}
\langle \, {J}_{\Lambda}, {J}_{\Omega} \,
\rangle_{\beta} =
c_{\Lambda}(\beta) \, \delta_{\Lambda \Omega} \,
,
\end{equation}
where $c_{\Lambda}(\beta)$ is some function of
$\beta$ (to be determined
in the next
section).
\end{theorem}

\section{Symmetryzing the
non-symmetric Jack polynomials in superspace}

For a composition
$\lambda$, the non-symmetric Jack polynomials,
$E_{\lambda}$, are the
unique polynomials in the variables
$z_1,\dots,z_N$
satisfying
\begin{equation} \label{defnonsym}
E_{\lambda} = z^{\lambda}
+ \sum_{\mu < \lambda} c_{\lambda
\mu}(\beta) z^{\mu} \, ,
\qquad
{\rm and} \qquad \langle \, E_{\lambda},E_{\mu}
\,
\rangle_{\beta} \propto \delta_{\lambda \mu}.
\end{equation}
The
non-symmetric Jack polynomial
$E_{\lambda}$ is an eigenfunction of
the Dunkl operators,
\begin{equation}
{\mathcal D}_i E_{\lambda} =
\underline{ \lambda_i} E_{\lambda} \, ,
\end{equation}
where the
eigenvalue $\underline{\lambda}_i$ is given in
(\ref{lsous}).
This
property characterizes $E_{\lambda}$ uniquely
\cite{Opdam}.

Using the orthogonality of the non-symmetric Jack
polynomials,
the polynomials $E_{\lambda,I}
=\theta_{I} E_{\lambda}$
are immediately seen to form an orthogonal basis
of $P$.  These
polynomials are in fact the unique common eigenfunctions
of the
operators ${\mathcal D}_i$ and $\theta_i \frac{\d}{ \d
\theta_i}$,
for $i=1,\dots,N$.

A basis of the space of
$S_N$-symmetric polynomials in the variables
$z_1,\dots,z_N$ is given
by the Jack polynomials.  The following formula is
known \cite{Opdam,
sahi}
\begin{equation}
J_{\lambda^+} \propto \sum_{w \in S_N} K_{w}
E_{\lambda^+} \, ,
\end{equation}
where $J_{\lambda^+}$ is the Jack
polynomial indexed by the
partition $\lambda^+$.  On the operatorial
side, the Jack polynomials are the
unique common eigenfunctions of
the operators ${\mathcal H}_n=
\sum_{i=1}^N {\mathcal D}_i^n$
($n=1,2,\dots$).

Other polynomials obtained from the non-symmetric
Jack polynomials have been
studied in \cite{Baker, Baker2}.  We are
particularly interested in those obtained by antisymmetrizing
the first
$m$ variables and symmetrizing the remaining ones.
Namely, given
partitions
$\lambda$ and $\mu$ with $m$ and $N-m$
parts respectively,
let\footnote{In \cite{Baker}, the antisymmetrization (resp. symmetrization)
is performed
on the last $m$ variables (resp. first $N-m$ variables).
This does not affect in any meaningful way the properties of these
polynomials.  For instance, formula \eqref{normBF} can be extracted easily
from a similar formula of \cite{Baker}.
}
\begin{equation}\label{lesS}
S_{(\lambda,\mu)}
=
\frac{(-1)^{m(m-1)/2}}{f_{\mu}} \sum_{\sigma \in
S_{m}}
\sum_{w \in S_{m^c}} K_{\sigma}\,  K_{w} \,
(-1)^{\ell(\sigma)}
      E_{(\lambda^R,\mu^R)}
\, ,
\end{equation}
where $E_{(\lambda^R,\mu^R)}$ is the non-symmetric
Jack polynomial
indexed by the
concatenation of the compositions
$\lambda^R$ (recall that this is
the partition $\lambda$
in reversed
order) and
$\mu^R$ (that is, the adjunction of the entries of $\mu^R$
to the
right of those of
$\lambda^R$ without rearrangement), and
where $S_{m^c}$ stands for
the permutations of
$\{m+1,\dots,N \}$
(or the permutations of $S_N$ that leave
$1,\dots,m$
fixed).

\begin{property} \cite{Baker}.
The polynomials
$S_{(\lambda,\mu)}$  are
orthogonal with respect to the scalar
product $\langle\, , \,
\rangle_\beta$, the norm
being given
explicitly by
\begin{equation}\label{normBF}
\langle \,
S_{(\lambda,\mu)},S_{(\lambda,\mu)} \,
\rangle_{\beta} =
\frac{m!
(N-m)!}{f_{\mu}}\;
\frac{d'_{(\lambda^R, \mu)}d_{(\lambda^R, \mu^R)}}
{d'_{(\lambda^R,\mu^R)}d_{(\lambda,\mu^R)}}
\langle
\, E_{(\lambda^R,\mu^R)},E_{(\lambda^R, \mu^R)}\, \rangle_{\beta} \, ,
\end{equation}
where for a composition $\gamma$ (in this case given by the
concatenation of two
compositions)
\begin{eqnarray}
d_\gamma &=& \prod_{(i,j)\in\gamma}[a(i,j)+1+\beta(\, \l(i,j)+1)] \nonumber \\
d'_\gamma &=& \prod_{(i,j)\in\gamma}[a(i,j)+1+\beta\, \l(i,j)]
\end{eqnarray}
with
\begin{eqnarray}
a(i,j) &=& \gamma_i-j \nonumber \\
l(i,j) &=& \# \{k=1,...,i-1 \, | \, j \leq \gamma_k+1\leq \gamma_i \}
+ \# \{k=i+1,...,N \, |\,  j \leq\gamma_k \leq \gamma_i \}
\end{eqnarray}
and
\begin{equation}
\langle \, E_{\gamma},E_{\gamma} \, \rangle_{\beta}= \prod_{1\leq
i<j}^N\prod_{p=0}^{\beta-1}\left(\frac{
{\underline\gamma_j}-{\underline\gamma_i}
+p} { {\underline\gamma_j}-{\underline\gamma_i}
-p-1}\right)^{\epsilon({\underline\gamma_j}-{\underline\gamma_i})} \, ,
\end{equation}
where
$\epsilon(x)=1$ if $x>0$ and $-1$ otherwise.
\end{property}

We now build a basis of $P^{S_N}$ from the non-symmetric
Jack polynomials.
\begin{definition} \label{defJ} Given a superpartition $\Lambda=(\La^a;\La^s)$,
\begin{equation}
\tilde {J}_{\Lambda} = \frac{(-1)^{(m)(m-1)/2}}{f_{\La^s}}
\sum_{w \in S_N} {\mathcal K}_{w} \, \theta_{\{1,\dots,m\}} \,
E_{((\La^a)^R,(\La^s)^R)} \, .
\end{equation}
\end{definition}
\begin{proposition} \label{propS}
We have
\begin{equation}
\tilde {J}_{\Lambda} = \sum_{w \in S_N/(S_{m} \times S_{m^c})}
\K_w \, \theta_{\{1,\dots,m \}} \, S_{(\La^a,\La^s)} \, .
\end{equation}
\end{proposition}
\n {\it Proof:}
From the definition of $\tilde {J}_{\Lambda}$, we get
%\begin{equation}
\begin{eqnarray}
\tilde {J}_{\Lambda} &=& \frac{(-1)^{(m)(m-1)/2}}{f_{\Lambda^s}}
\sum_{w \in S_N/(S_{m}\times S_{m^c})}
       {\mathcal K}_{w}  \sum_{\sigma \in S_{m}}
       {\mathcal K}_{\sigma} \sum_{\rho \in S_{m^c}}
       {\mathcal K}_{\rho}
\, \theta_{\{1,\dots,m \}} \,
E_{((\La^a)^R,(\La^s)^R)}  \nonumber  \\
       &=&
\! \! \! \sum_{w \in S_N/(S_{m}\times S_{m^c})}
       \! \! \! {\mathcal K}_{w}\,
       \theta_{\{1,\dots,m \}} \left(
\frac{(-1)^{(m)(m-1)/2}}{f_{\La^s}} \sum_{\begin{subarray}{c}
\sigma \in S_{m} \\
\rho \in S_{m^c}
\end{subarray}
}
\! \! \!  {\mathcal K}_{\sigma}   {\mathcal K}_{\rho}  \, (-1)^{\ell(\sigma)}
E_{((\La^a)^R,(\La^s)^R)}\right) \, ,
\end{eqnarray}
%\end{equation}
which gives the desired result from the definition of $S_{(\La^a,\La^s)}$.
\hfill $\square$

Note that the left coset representatives of
$ S_N/(S_{m}\times S_{m^c})$ can be
described as
\begin{equation}
\left\{\sum_{k=0}^{{\rm min}(m,N-m)} \sum_{1\leq i_1<i_2< \cdots <
i_k\leq m\atop m+1\leq
j_1<j_2\cdots <j_k\leq N}\K_{i_1,j_i}\cdots
\K_{i_k,j_k} \right\} \, ,
\end{equation}
with the understanding that when $k=0$, the product of $\K$ factors
reduces to the
identity.

\begin{proposition} \label{propmonotri} We have
\begin{equation}
\tilde {J}_{\Lambda} = {m}_{\Lambda}+ \sum_{\Omega < \Lambda}
a_{\Lambda \Omega}(\beta) \, m_{\Omega} \, ,
\end{equation}
that is, $\tilde {J}_{\Lambda}$ is monic and
triangularly related (with respect to the Bruhat ordering on superpartitions)
 to the monomial superfunction basis.
\end{proposition}
\n {\it Proof:} \quad  The monicity of $\tilde {J}_{\Lambda}$, given
the monicity of $S_{(\lambda,\mu)}$ \cite{Baker},
follows from Proposition~\ref{propS}.
Now, let $\Lambda^R$ stand for $((\Lambda^a)^R,(\Lambda^s)^R)$.
From Definition~\ref{defJ} and \eqref{defnonsym}, the coefficient of $\theta_1 \cdots \theta_m$ in
$\tilde {J}_{\Lambda}$ is
\begin{eqnarray}
\tilde {J}_{\Lambda} \Big |_{\theta_1 \cdots \theta_m} &=&
\frac{(-1)^{m(m-1)/2}}{f_{\Lambda^s}} \sum_{\sigma \in S_m} \sum_{w \in
S_{m^c}} \K_\sigma \K_w \, (-1)^{\ell(\sigma)} E_{\Lambda^R}
\nonumber \\
&=&
\frac{(-1)^{m(m-1)/2}}{f_{\Lambda^s}} \sum_{\sigma \in S_m} \sum_{w \in
S_{m^c}} \K_\sigma \K_w \, (-1)^{\ell(\sigma)} \sum_{\Omega \leq \Lambda^R}
c_{\Lambda \Omega}(\beta) \, z^\Omega \, ,
\end{eqnarray}
where in the last equality, the ordering is on compositions.
To obtain the monomial superfunctions that appear in the expansion of
$\tilde {J}_{\Lambda}$, we must simply select the superpartitions that
arise as powers of $z$ in this last equality.
 Because of the nature of the sums over
$\sigma$ and $w$, the only superpartition $\Gamma$ such that
$\Gamma^*=\Lambda^*$ that can arise is $\Lambda$.  Finally, since any
rearrangement $\Gamma$ of a composition $\Omega$ such
that $\Omega< \Lambda^R$ and $\Omega^* \neq \Lambda^*$ is
also such that $\Gamma< \Lambda$, the proposition
is seen to hold.
\hfill
$\square$

The orthogonality of the $\tilde {J}_{\Lambda}$'s is almost
immediate from Proposition~\ref{propS}.
\begin{proposition} We have
\begin{equation}
\langle \, \tilde {J}_{\Lambda}, \tilde {J}_{\Omega} \, \rangle_{\beta}
= \delta_{\Lambda \Omega} \frac{N!}{m!(N-m)!}
\langle \,  {S}_{(\La^a,\La^s)}, {S}_{(\La^a,\La^s)} \, \rangle_{\beta}
\, ,
\end{equation}
where $\langle \,
{S}_{(\La^a,\La^s)}, {S}_{(\La^a,\La^s)} \, \rangle_{\beta}$
is given explicitely in \eqref{normBF}.
\end{proposition}
\n {\it Proof:} \quad Using Proposition~\ref{propS}, we have
\begin{equation}
\langle \, \tilde {J}_{\Lambda}, \tilde {J}_{\Omega} \,  \rangle_{\beta}
= \Bigl \langle \, \sum_{w \in G}
\K_w \, \theta_{\{1,\dots,m \}} \, S_{(\La^a,\La^s)} \, ,
\sum_{\sigma \in G}
\K_\sigma \, \theta_{\{1,\dots,m \}} \, S_{(\Omega^a,\Omega^s)}
\Bigr \rangle_{\beta}
\, ,
\end{equation}
where $G$ is a set of left coset representatives
of $S_N/(S_{m} \times S_{m^c})$.
Since $\K_{\sigma}^{\dagger}=\K_{\sigma^{-1}}$, this gives
\begin{equation}
\langle \, \tilde {J}_{\Lambda}, \tilde {J}_{\Omega} \, \rangle_{\beta}
= \Bigl \langle \, \sum_{w,\sigma \in G}
\K_{\sigma^{-1}} \K_w \, \theta_{\{1,\dots,m \}} \, S_{(\La^a,\La^s)} \, ,
       \theta_{\{1,\dots,m \}} \, S_{(\Omega^a,\Omega^s)} \Bigr
\rangle_{\beta}
\, .
\end{equation}
Now, $\K_{\sigma^{-1}} \K_w$ must belong to $S_m \times S_{m^c}$
for the product
of $\theta$'s
to be the same on both sides.
  Therefore, since this only occurs for $w=\sigma$,
\begin{equation}
\langle \, \tilde {J}_{\Lambda}, \tilde {J}_{\Omega} \, \rangle_{\beta}
=\# G \,  \langle \,   S_{(\La^a,\La^s)} \, ,
       S_{(\Omega^a,\Omega^s)} \, \rangle_{\beta}
\, ,
\end{equation}
which proves the proposition since the cardinality of $G$ is $N!/\bigl(m! \,
(N-m)!\bigr)$.
\hfill $\square$

\medskip

We now make the connection with the family
$\{ {J}_{\Lambda}\}_{\Lambda}$ introduced before.
\begin{theorem} \label{jjbar} For any superpartition $\Lambda$, we have
${J}_{\Lambda}=\tilde {J}_{\Lambda}$.
\end{theorem}
\n {\it Proof:} \quad  Both families are triangular with respect to the
same ordering (namely, the  Bruhat ordering on compositions) when expanded
in the supermonomial basis.  Since the Gram-Schmidt
orthonormalization procedure
ensures that there exists at most one orthonormal family with such
triangularity, the
two families can only differ by a constant.  The
theorem then follows
from the monicity of both families.
\hfill
$\square$

\section{Conclusion}

This work has many natural
generalizations. The most direct one is
to consider the rational
counterpart of
the above results.  The
orthogonal eigenfunctions of
the supersymmetric rational CMS model
can be obtained in
a rather
direct way using the present results and the remarkable
relation,
preserved in the supersymmetric case,  that exists between
the
eigenfunctions of the
trigonometric and rational
models
\cite{Sogo}. This leads to a closed form expression (a
determinant of
determinants) for the orthogonal generalized Hermite
(or Hi-Jack)
superpolynomials.  These results will be presented
in
\cite{DLM3}.

      Another rather immediate line of
generalizations would be to examine
the supersymmetric
extension of
the r/tCMS models associated to root systems
of any type. To
find the
corresponding orthogonal superpolynomials, one would proceed
as
follows:
take the Dunkl
operator of the corresponding exchange
version of the r/tCMS model of
interest, look for
their non-symmetric
eigenfunctions, dress them with a fermionic
monomial prefactor
and
symmetrize the result with respect to both types of variables.
For instance,
the generalized  Jacobi and Laguerre polynomials in
superspace could
be constructed in this way. The resulting
superpolynomial would be a linear
combination (with $\theta_I$
coefficients) of the corresponding version
(Jacobi or
Laguerre) of the generalized  polynomials with mixed
symmetry
\cite{Baker2}. Note, however, that the norm of these
special
polynomials has not yet been computed.  The conserved charges
of the
model would
be
constructed exactly as in the present case, by
symmetryzing the Dunkl
operators raised to the
$n$-th power and
multiplied by a fermionic prefactor.

Along these lines, the
formulation  of the supersymmetric
extension of the elliptic (eCMS)
model appears to be
    direct.
The elliptic Dunkl operators are given
in \cite{BFV}.  Their
supersymmetric lift is immediate, leading
directly to an expression for
the Hamiltonian of the
seCMS model.
Again, the general form of the
charges is bound to be similar to the
one found in this article.
However, in this case,
little is known
about the eigenfunctions.

      The $q$ deformation of the Jack
polynomials are the Macdonald polynomials
\cite{Mac}, eigenfunctions
(up to a conjugation)
of the Ruijsenaars-Schneider model
\cite{Ruij},
a relativistic version of the CMS model.  Again, there exist
$q$
analogues of the
Dunkl operators, giving a natural road for the
formulation of the
supersymmetric Ruijsenaars-Schneider model.
Moreover, the
nonsymmetric Macdonald
polynomials being known, they
can be lifted to orthogonal
Macdonald superpolynomials.

      Further avenues regarding future works concern
the study of
properties of the Jack
superpolynomials $J_{\La}$
themselves. On that matter, we already have strong
indications that
these objects have
       rather remarkable
properties.  In
particular,  the Pieri formulas appear  to be
rather nice. Moreover,
exploratory analyses indicate that $J_\La$
products have a
combinatorial interpretation in terms of novel
types of supertableaux.
Finally, a natural problem that should not
be out at reach at this
stage is
    working out the superspace extension
     of the operator
construction of \cite{LV}.

\vskip0.3cm
\noindent {\bf
ACKNOWLEDGMENTS}

This work was  supported by
NSERC. L.L. wishes to
thank Luc Vinet for his
support, and P.D. is grateful to the
Fondation
J.A Vincent
for a
student
fellowship.

\begin{appendix}

\section{Examples of Jack
superpolynomials}

In this appendix, we present a detailed
calculation of one
Jack polynomial in superspace via determinantal
formulas. Then, we
give explicitly all orthogonal surperpolynomials
of
degrees not larger than 3.

Let $\Lambda=(3,1;0)$.  It has
weight
$4$ and lies in the 2-fermion sector.  The explicit action of
the
conserved operator $\I_1$ in the
space
$\mbox{Span}\{\mathcal{J}_\Omega\}_{\Omega<_t\Lambda}$ is
obtained
from Lemma~\ref{lem5}:
\begin{eqnarray}
\I_1\, \sj{3,1;0}&=&(N-1)!\, (4-2\beta)\, \sj{3,1;0}+(N-1)!\, \beta\,
\sj{3,0;1}-
(N-1)!\, \beta\, \sj{1,0;3}\,
, \cr
   \I_1\, \sj{3,0;1}&=&(N-1)!\, (3-3\beta)\, \sj{3,0;1}+(N-1)!\,
\beta\, \sj{1,0;3}\, ,\cr
\I_1\, \sj{1,0;3}&=&(N-1)!\, (1-4\beta)\, \sj{1,0;3}\, .
\end{eqnarray}
Theorem~\ref{deter} allows to express the orthogonal
superpolynomial $J_{(3,1;0)}$ as the following determinant:
\begin{eqnarray}
{J}_{(3,1;0)}&=&\frac{1}{(N-1)!\, (3+2\beta)(N-1)!\,
(1+\beta)}\left|\begin{matrix}
    \sj{1,0;3} & -(N-1)!\, (3+2\beta) &0\\
   \sj{3,0;1} & (N-1)!\, \beta &-(N-1)!\, (1+\beta)\\
   \sj{3,1;0}&-(N-1)!\, \beta&(N-1)!\, \beta
\end{matrix}\right|\cr
&=&\sj{3,1;0}+\frac{\beta}{1+\beta}\sj{3,0;1}-\frac{\beta}{(3+2\beta)(1+\beta)}\sj{1,0;3}\,
.
\end{eqnarray}

We now want the monomial decomposition of the previous result. For
this, we must determine the action of the Hamiltonian $\H_2$ in the
space $\mbox{Span}\{{m}_\Omega\}_{\Omega<_h\Lambda}$. Using
Theorem~2 in Ref.~\cite{DLM2}, the action of $\H_2$ on the monomial
$\sm{3,1;0}$ is given by
\begin{equation}
\H_2 \, \sm{3,1;0 }= (10-6\beta+4 N \beta
)\, \sm{3,1;0}+4\beta\, \sm{2,1;1}\, ,
\end{equation}
The monomial $\sm{2,1;1}$ is itself an eigenfunction of the
Hamiltonian, i.e.,
\begin{equation}
\H_2 \, \sm{2,1;1 }= (6-10\beta+4 N \beta)\, \sm{2,1;1}\, .
\end{equation}
Again, Theorem~\ref{deter} yields the non-orthogonal
superpolynomial $\mathcal{J}_{(3,1;0)}$ as a
determinant:
\begin{eqnarray}
\mathcal{J}_{(3,1;0)}&=&\frac{1}{(4\beta)(-4-4\beta)}\left|\begin{matrix}
    \sm{2,1;1} & -4-4\beta \\
   \sm{2,1;0} & 4\beta
\end{matrix}\right|\cr
&=&\sm{3,1;0}+\frac{\beta}{1+\beta}\sm{2,1;1}
\end{eqnarray}
Proceeding in the same way, we get:
{\small
\begin{eqnarray}
  \mathcal{J}_{(3,0;1)}
&=&\frac{1}{(-6-10\beta)(-4-4\beta)^3(-2-2\beta)}\left|\begin{matrix}
    \sm{1,0;1^3}  &-6-10\beta&0   &0      &0      &0\cr
      \sm{1,0;2,1}  &12\beta&-4-4\beta&0    &0    &0\cr
    \sm{2,0;1^2}  &6\beta&0       &-4-4\beta&0    &0\cr
    \sm{2,1;1}    &0      &0      &0      &-4-4\beta&0\cr
    \sm{2,0;2}    &0      &2\beta &4\beta &2\beta &-2-2\beta\cr
    \sm{3,0;1}    &0      &2\beta &8\beta &2\beta &2\beta\cr
\end{matrix}\right|\cr
&=&\sm{3,0;1}+\frac{\beta}{1+\beta}\sm{2,0;2}+\frac{\beta(1+2\beta)}{2(1+\beta)^2}\sm{2,1;1}
  \cr
&& \qquad \qquad
+\frac{\beta(2+3\beta)}{(1+\beta)^2}\sm{2,0;1^2}
  +\frac{\beta(1+2\beta)}{2(1+\beta)^2}\sm{1,0;2,1}+\frac{3
\beta^2}{(1+\beta)^2}\sm{1,0;1^3}\end{eqnarray}
}
  and
{\small
\begin{eqnarray}
\mathcal{J}_{(1,0;3)}&=&\frac{1}{(-6-10\beta)(-4-4\beta)^3(-2-2\beta)}\left|\begin{matrix}
    \sm{1,0;1^3}  &-6-10\beta&0   &0      &0      &0\cr
      \sm{1,0;2,1}  &12\beta&-4-4\beta&0    &0    &0\cr
    \sm{2,0;1^2}  &6\beta&0       &-4-4\beta&0    &0\cr
    \sm{2,1;1}    &0      &0      &0      &-4-4\beta&0\cr
    \sm{2,0;2}    &0      &2\beta &4\beta &2\beta &-2-2\beta\cr
    \sm{1,0;3}    &0      &6\beta &0 &-2\beta &2\beta\cr
\end{matrix}\right|\cr
&=&\sm{3,0;1}+\frac{\beta}{1+\beta}\sm{2,0;2}-\frac{\beta}{2(1+\beta)^2}\sm{2,1;1}
\cr
   & & \qquad \qquad +\frac{\beta^2}{(1+\beta)^2}\sm{2,0;1^2}
+\frac{\beta(3+4\beta)}{2(1+\beta)^2}\sm{1,0;2,1}+\frac{3
\beta^2}{(1+\beta)^2}\sm{1,0;1^3}\, .\end{eqnarray}
}
Consequently,
the orthogonal Jack superpolynomial $J_{(3,1;0)}$ is written in the
monomial basis as:
\begin{eqnarray}
J_{(3,1;0)}&=&\sm{3,1;0}+\frac{\beta}{1+\beta}\sm{3,0;1}-\frac{\beta}{(1+\beta)(3+2\beta)}\sm{1,0;3}
\cr & &+\frac{2\beta^2}{(1+\beta)(3+2\beta)}\sm{2,0;2}
+\frac{\beta(3+4\beta)}{(1+\beta)(3+2\beta)}\sm{2,1;1} \cr
   &
&+\frac{6\beta^2}
{(1+\beta)(3+2\beta)}\sm{2,0;1^2}+\frac{2\beta^3}
{(1+\beta)^2(3+2\beta)}\sm{1,0;2,1}
+\frac{6\beta^3}
{(1+\beta)^2(3+2\beta)}\sm{1,0;1^3}
.\end{eqnarray}
%%** Luc: le dernier terme est bien la

Finally, we give the simplest Jack
superpolynomials explicitly in Tables \ref{tabjack1} and \ref{tabjack2}.
The polynomials have degrees less than 3 and 4 in $\theta$ and $z$
respectively. In the first table, the non-orthogonal
eigenfunctions $\mathcal{J}_\Lambda$ are written in  the monomial
basis $\{m_\Omega\}_{\Omega<_h \Lambda}$.  The second table
presents the orthogonal Jack polynomials in the non-orthogonal
basis $\{\mathcal{J}_\Omega\}_{\Omega<_t \Lambda}$.  We stress
that these expressions do not depend on the number of variables
$N$.

   \begin{table} \caption{Non-orthogonal
superpolynomials $\mathcal{J}_\Lambda$ of weight $|\Lambda|\leq 3$
}
\begin{center}
\label{tabjack1}
\begin{tabular}{l|l} \hline
   Superpartition
    & \qquad \qquad \qquad \qquad \qquad \qquad Superpolynomial \\
\qquad\,\,$\Lambda$& \qquad \qquad \qquad \qquad \qquad \qquad
\quad$\mathcal{J}_{\Lambda}(z,\theta;1/\beta)$ \\
\hline\\
$(0)$&$\sm{0}$\\
$(1)$&$\sm{1}$\\
$(1^2)$&$\sm{1^2}$\\
$(2)$&$\sm{2}+\frac{2\beta}{1+\beta}\sm{1^2}$\\
$(1^3)$&$\sm{1^3}$\\
$(2,1)$&$\sm{2,1}+\frac{6\beta}{1+2\beta}\sm{1^3}$\\
$(3)$&$\sm{3}+\frac{3\beta}{2+\beta}\sm{2,1}+\frac{6\beta^2}{(1+\beta)(2+\beta)}\sm{1^3}$\\
&\\
$(0;0)$ &$\sm{0;0} $\\
   $(0;1)$ & $ \sm{0;1}$\\
   $(1;0) $& $ \sm{1;0}$ \\
    $(0;1^2)$ &$ \sm{0;1^2}$ \\
    $(1;1)$& $ \sm{1;1}$ \\
    $(0;2)$ & $
\sm{0;2}+\frac{\beta}{1+\beta}\sm{1;1}+\frac{2\beta}{1+\beta}\sm{0;1^2}$\\
    $(2;0) $& $ \sm{2;0}+\frac{\beta}{1+\beta}\sm{1;1} $\\
     $(0;1^3) $& $ \sm{0;1^3} $ \\
   $(1;1^2)$ & $ \sm{1;1^2}$\\
   $(0;2,1) $& $
m_{(0;2,1)}+\frac{2\beta}{1+2\beta}m_{(1;1^2)}+\frac{6\beta}{1+2\beta}m_{(0;1^3)}$

\\
$(1;2)$ & $  m_{(1;2)}+\frac{2\beta}{1+2\beta}m_{(1;1^2)}$ \\
   $(2;1) $& $  m_{(2;1)}+\frac{2 \beta}{1+2\beta}m_{(1;1^2)}$ \\
$(0;3)$ & $\sm{0;3}+\frac{\beta}{2+\beta}
\sm{2;1}+\frac{2\beta}{2+\beta} \sm{1;2}+\frac{3\beta}{2+\beta}
\sm{0;2,1}
+\frac{4\beta^2}{(1+\beta)(2+\beta)}
\sm{1;1^2}+\frac{6\beta^2}{(1+\beta)(2+\beta)} \sm{0;1^3}$ \\
   $(3;0)$ & $\sm{3;0}+\frac{2\beta}{2+\beta}
\sm{2;1}+\frac{\beta}{2+\beta}
\sm{1;2}+\frac{2\beta^2}{(1+\beta)(2+\beta)} \sm{1;1^2}$ \\

   &\\
    $(1,0;0) $& $ \sm{1,0;0}$ \\
    $(1,0;1)$ & $ \sm{1,0;1}$ \\
    $(2,0;0)$ & $ \sm{2,0;0}+\frac{\beta}{1+\beta}\sm{1,0;1}$ \\
     $(1,0;1^2)$ & $ \sm{1,0;1^2}$ \\
     $(1,0;2)$ & $\sm{1,0;2}+\frac{2 \beta}{1+2\beta}\sm{1,0;1^2}  $ \\
$(2,0;1) $& $\sm{2,0;1}+\frac{2 \beta}{1+2\beta}\sm{1,0;1^2} $ \\
$(2,1;0) $& $ \sm{2,1;0}$ \\
   $(3,0;0)$ & $
\sm{3,0;0}+\frac{\beta}{2+\beta}\sm{2,1;0}+\frac{2\beta}{2+\beta}\sm{2,0;1}+\frac{\beta}{2+\beta}\sm{1,0;2}
+\frac{2\beta^2}{(1+\beta)(2+\beta)}\sm{1,0;1^2}$ \\
&\\ \hline \end{tabular}
\end{center}
   \end{table}

   \begin{table} \caption{Orthogonal Jack
superpolynomials ${J}_\Lambda$ of weight $|\Lambda|\leq 3$ }
\begin{center}
\label{tabjack2}
\begin{tabular}{l|l} \hline
   Superpartition
    & \qquad \qquad  Jack superpolynomial \\
\qquad\,\,$\Lambda$& \qquad \qquad \qquad \,\,
${J}_{\Lambda}(z,\theta;1/\beta)$ \\
\hline\\
$(0)$&$\sj{0}$\\
$(1)$&$\sj{1}$\\
$(1^2)$&$\sj{1^2}$\\
$(2)$&$\sj{2}$\\
$(1^3)$&$\sj{1^3}$\\
$(2,1)$&$\sj{2,1}$\\
$(3)$&$\sj{3}$\\
&\\
$(0;0)$ &$\sj{0;0} $\\
   $(0;1)$ & $ \sj{0;1}$\\
   $(1;0) $& $ \sj{1;0}+\frac{\beta}{1+\beta}\sj{0;1}$ \\
    $(0;1^2)$ &$ \sj{0;1^2}$ \\
    $(1;1)$& $ \sj{1;1}+\frac{2\beta}{1+2\beta}\sj{0;1^2}$ \\
    $(0;2)$ & $ \sj{0;2}$\\
    $(2;0) $& $ \sj{2;0}+\frac{\beta}{2+\beta}\sj{0;2} $\\
     $(0;1^3) $& $ \sj{0;1^3} $ \\
   $(1;1^2)$ & $ \sj{1;1^2}+\frac{3\beta}{1+3\beta}\sj{0;1^3}$\\
   $(0;2,1) $& $  \sj{0;2,1}$ \\
$(1;2)$ & $  \sj{1;2}+\frac{\beta}{1+\beta}\sj{0;2,1)}$ \\
   $(2;1) $& $  \sj{2;1}+\frac{\beta}{1+\beta}\sj{1;2}+\frac{
\beta(1+2\beta)}{2(1+\beta)^2}\sj{0;2,1}$ \\
$(0;3)$ & $\sj{0;3}$ \\
   $(3;0)$ & $\sj{3;0}+\frac{\beta}{3+\beta} \sj{0;3}$ \\

   &\\
    $(1,0;0) $ & $ \sj{1,0;0}$ \\
    $(1,0;1)$   & $ \sj{1,0;1}$ \\
    $(2,0;0)$   & $ \sj{2,0;0}$ \\
    $(1,0;1^2)$ &$\sj{1,0;1^2}$\\
     $(1,0;2)$  & $\sj{1,0;2} $ \\
$(2,0;1) $& $\sj{2,0;1}+\frac{\beta}{1+\beta}\sj{1,0;2} $ \\
$(2,1;0) $& $
\sj{2,1;0}+\frac{\beta}{1+\beta}\sj{1,0;2}-\frac{\beta}{2(1+\beta)^2}\sj{1,0;2}$

\\
   $(3,0;0)$ & $ \sj{3,0;0}$ \\
&\\ \hline \end{tabular}
\end{center}
   \end{table}

\end{appendix}

\end{document}